\documentclass[conference]{IEEEtran}
\IEEEoverridecommandlockouts

\usepackage{cite}
\usepackage{amsmath,amssymb,amsfonts}
\usepackage{graphicx}
\usepackage{textcomp}
\usepackage{xcolor}
\def\BibTeX{{\rm B\kern-.05em{\sc i\kern-.025em b}\kern-.08em
    T\kern-.1667em\lower.7ex\hbox{E}\kern-.125emX}}
    
\usepackage{pdfpages}
\usepackage{changepage}
\usepackage{graphicx}
\usepackage{subcaption}
\usepackage{hyperref}
\usepackage{algpseudocode}
\usepackage{amsmath}
\usepackage{tabularray}
\usepackage[english]{babel}
\usepackage[nolist]{acronym}
\usepackage{pifont}
 \usepackage[disable]{todonotes}
\usepackage[table]{xcolor}
\usepackage{tabularx}
\usepackage{url}
\usepackage[mathscr]{euscript}
\usepackage{xspace}
\usepackage{makecell} 
\usepackage{booktabs}
\usepackage{enumitem}
\usepackage{multirow}
\usepackage[most]{tcolorbox}
\usepackage{balance}
\usepackage[nolist]{acronym}
\usepackage{svg}
\definecolor{darkyellow}{HTML}{ffda33}
\newcommand*{\emptycircle}{\begin{tikzpicture}[scale=0.1]
    \draw[black, thick, fill=white] (0,1) circle (1);
    \end{tikzpicture}\xspace}
\newcommand*{\halfcircle}{\begin{tikzpicture}[scale=0.1]
    \draw[black, thick] (0,1) circle (1);
    \fill[fill opacity=0.7,fill=black] (0,1) -- (90:1.9) arc (90:90-50*3.6:0.9) -- cycle;
    \end{tikzpicture}\xspace}
\newcommand*{\fullcircle}{\begin{tikzpicture}[scale=0.1]
    \draw[black, thick, fill=black] (0,1) circle (1);
    \end{tikzpicture}\xspace}

\definecolor{highlightcolor}{RGB}{184, 222, 162}   

\newcommand{\transcend}{$D_T$\xspace}
\newcommand{\apigraph}{$D_A$\xspace}

\newcommand{\transcendbold}{$\mathbf{D_T}$\xspace}
\newcommand{\apigraphbold}{$\mathbf{D_A}$\xspace}

\newcommand{\androzoo}{AndroZoo\xspace}
\newcommand{\longvt}{VirusTotal\xspace}
\newcommand{\shortvt}{VT\xspace}
\newcommand{\vtthreshold}{VTT\xspace}

\begin{acronym}
    \acro{ML}{Machine Learning}
    \acro{mlp}[MLP]{Multilayer Perceptron}
    \acro{dnn}[DNN]{Deep Neural Network}
    \acro{AUT}{Area Under Time}
    \acro{A-AUT}{Average Area Under Time}
    \acro{STD}{Standard Deviation}
    \acro{AV}{Anti-Virus}
    \acro{API}{Application Programming Interface}
    \acro{apk}[APK]{Android Application Package}
    \acro{vt}[VT]{VirusTotal}
    \acro{vtt}[VTT]{VirusTotal Threshold}
    \acro{az}[AZ]{AndroZoo}
    \acro{tpr}[TPR]{True Positive Rate}
    \acro{fpr}[FPR]{False Positive Rate}
    \acro{dtw}[DTW]{Dynamic Time Warping}
\end{acronym}

\newcommand{\drebin}{\textsc{Drebin}\xspace}
\newcommand{\deepdrebin}{\textsc{DeepDrebin}\xspace}
\newcommand{\malscan}{\textsc{MalScan}\xspace}
\newcommand{\ramda}{\textsc{Ramda}\xspace}
\newcommand{\HCC}{\textsc{Hcc}\xspace}

\newcommand{\dada}{\textsc{DaDa}\xspace}
\newcommand{\stas}{\textsc{Stas}\xspace}
\newcommand{\hypercube}{\textsc{Hypercube}\xspace}
\newcommand{\googleplay}{GooglePlay\xspace}

\newcommand{\vtdate}{\textit{\shortvt~first\_submission\_date}\xspace}
\newcommand{\vtdates}{\textit{VT\_first\_dates}\xspace}
\newcommand{\dexdate}{\textit{dex\_date}\xspace}
\newcommand{\dexdates}{\textit{dex\_dates}\xspace}
\newcommand{\uploaddate}{\textit{upload\_date}\xspace}
\newcommand{\uploaddates}{\textit{upload\_dates}\xspace}

\newcommand{\crawldates}{\textit{crawl\_dates}\xspace}

\newcommand{\fone}{$F_1$-Score\xspace}
\newcommand{\fonebold}{$\mathbf{F_1}$-Score\xspace}

\newcommand{\tesseract}{\textsc{Tesseract}\xspace}

\newcommand{\factorTimestampName}{Timestamp Type\xspace}
\newcommand{\factorMarketsName}{App Markets\xspace}
\newcommand{\factorVTTName}{\longvt Threshold\xspace}
\newcommand{\factorSamplingName}{Dataset Size\xspace}
\newcommand{\factorTemporalLuckName}{Temporal Luck\xspace}

\newcommand{\daniel}[1]{\todo[inline,color=yellow]{\small Daniel: #1}}

\newcommand{\hypothesisbox}[2]{
\begin{tcolorbox}[
    colback=yellow!10,
    colframe=yellow!50,
    coltitle=black,
    title=#1,
    fonttitle=\bfseries,
    boxrule=1mm, 
    arc=4mm,
    left=2mm,
    right=2mm,
    top=2mm, 
    bottom=2mm,
    sharp corners=downhill,
    width=\linewidth,
]
#2
\end{tcolorbox}
}

\newcommand{\recommendationboxnew}[2]{\begin{tcolorbox}[
    colback=blue!10, 
    colframe=blue!30,
    coltitle=black,
    title=#1,
    fonttitle=\bfseries,
    boxrule=1mm, 
    arc=4mm,
    left=2mm,
    right=2mm,
    top=2mm, 
    bottom=2mm,
    sharp corners=downhill,
    width=\linewidth,]
#2
\end{tcolorbox}
}

\newcommand{\eg}{e.g.,\xspace} 
\newcommand{\ie}{i.e.,\xspace} 

\renewcommand{\paragraph}[1]{{\vskip 4pt \noindent\textbf{#1.} }}

\def\Snospace~{\S{}}

\begin{document}

\title{Beyond the \tesseract:\\Trustworthy Dataset Curation for Sound Evaluations of Android Malware Classifiers}

\author{
    \IEEEauthorblockN{Theo Chow$^{\dagger\ddagger}$, Mario D'Onghia$^{\ddagger}$, Lorenz Linhardt$^{\star\circ}$, Zeliang Kan$^{\dagger\diamond\footnotemark{1}}$,\\Daniel Arp$^{\mathsection}$, Lorenzo Cavallaro$^{\ddagger}$, Fabio Pierazzi$^{\ddagger}$}
    \IEEEauthorblockA{$^{\dagger}$King's College London, 
    $^{\star}$Technische Universit\"at Berlin,
    \\
    $^{\circ}$BIFOLD – Berlin Institute for the Foundations of Learning and Data,
    $^{\diamond}$HiddenLayer, Inc., 
    \\     
    $^{\mathsection}$Technische Universit\"at Wien, $^{\ddagger}$University College London}
}

\maketitle
\footnotetext[1]{Zeliang Kan contributed to this work during his Ph.D. studies at KCL.}

\begin{abstract}

The reliability of machine learning critically depends on dataset quality. While machine learning applied to computer vision and natural language processing benefits from high-quality benchmark datasets, cyber security often falls behind, as quality ties to the ability of accessing hard-to-obtain realistic data that may evolve over time. Android is, however, positioned uniquely in this ecosystem due to AndroZoo and other sources, which provide large-scale, continuously updated, and timestamped repositories of benign and malicious apps. 

Since their release, such data sources provided access to populations of Android apps that researchers can sample from to evaluate learning-based methods in realistic settings, i.e., over temporal frames to account for app evolution (natural distribution shift) and test datasets that reflect in-the-wild class ratios. Surprisingly, we observe that despite this abundance of data, performance discrepancies of learning-based Android malware detectors still persist even after satisfying such realistic requirements, which challenges our ability to understand what the state of the art in this field is. In this work, we identify five novel factors that influence such discrepancies: we show how such factors have been largely overlooked and the impact they have on providing sound evaluations. Our findings and recommendations help define a methodology for curating trustworthy datasets towards sound evaluations of Android malware classifiers.

\end{abstract}

\section{Introduction}

The foundation of trustworthy \ac{ML} research lies in the datasets used for evaluation. To ensure that results generalize beyond controlled lab-only settings, datasets must faithfully capture the distributions, dynamics, and challenges of the real world. However, unlike computer vision and natural language processing, where community benchmarks such as ImageNet~\cite{deng2009imagenet} and MMLU~\cite{hendrycks2020measuring} (or datasets in repositories such as UCI~\cite{uciurl} and Kaggle~\cite{kaggle}) provide stable and widely accepted testbeds, the Android malware domain lacks reliable and standardized benchmarks. This absence is primarily due to both restricted access to real-world attack data~\cite{sommer2010outside} and the highly non-stationary nature of adversarial environments~\cite{pendlebury2019tesseract}, in which attackers' strategies evolve continuously, hindering the curation of static benchmarks. The introduction of large-scale collections of benign or malicious \acp{apk} such as \androzoo~\cite{allix2016androzoo} and VirusShare~\cite{virusshare} has partially addressed this gap by enabling researchers to curate time-aware Android malware datasets in accordance to the best practices and guidelines introduced in \tesseract~\cite{pendlebury2019tesseract} and other works on recommendations to apply \ac{ML} to cybersecurity~\cite{dos_and_donts, sommer2010outside, flood2024bad}. However, we show that these measures alone are insufficient to guarantee reliable and unbiased assessments.

\begin{figure*}[t]
    \centering
    \begin{minipage}{0.19\linewidth}
        \includegraphics[width=\linewidth]{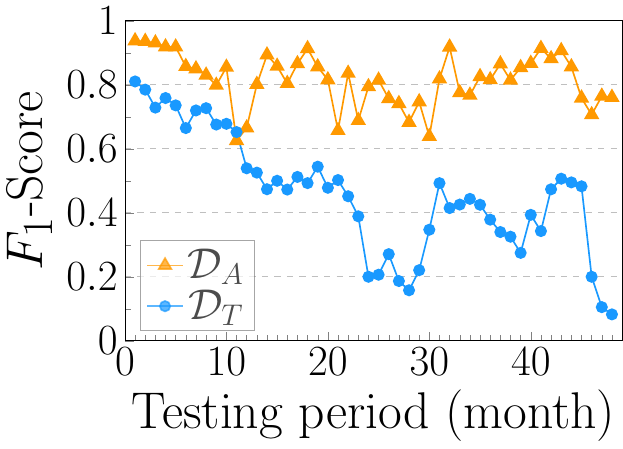}
        \subcaption{\drebin~\cite{arp2014drebin}}
    \end{minipage}
    \hfill
    \begin{minipage}{0.19\linewidth}
        \centering
        \includegraphics[width=\linewidth]{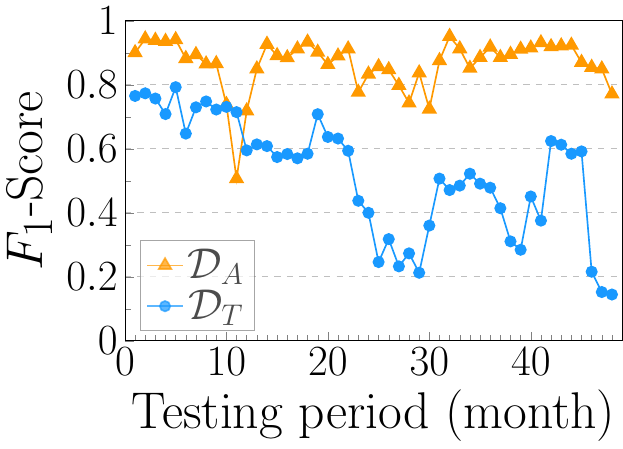}
        \subcaption{\deepdrebin~\cite{deepdrebin}}
    \end{minipage}
    \hfill
    \begin{minipage}{0.19\linewidth}
        \centering
        \includegraphics[width=\linewidth]{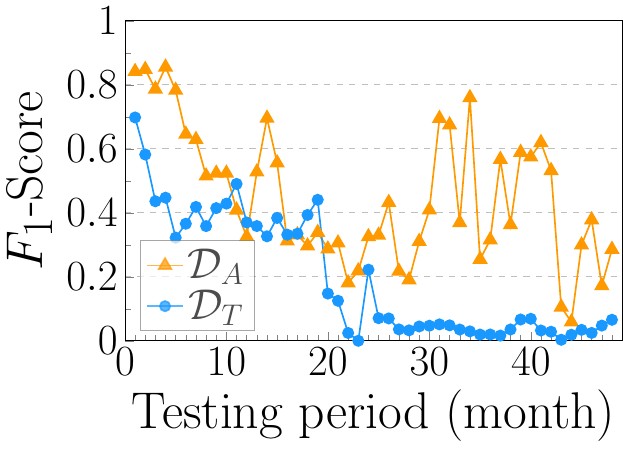}
        \subcaption{\malscan~\cite{malscan_paper}}
    \end{minipage}
    \hfill
    \begin{minipage}{0.19\linewidth}
        \centering
        \includegraphics[width=\linewidth]{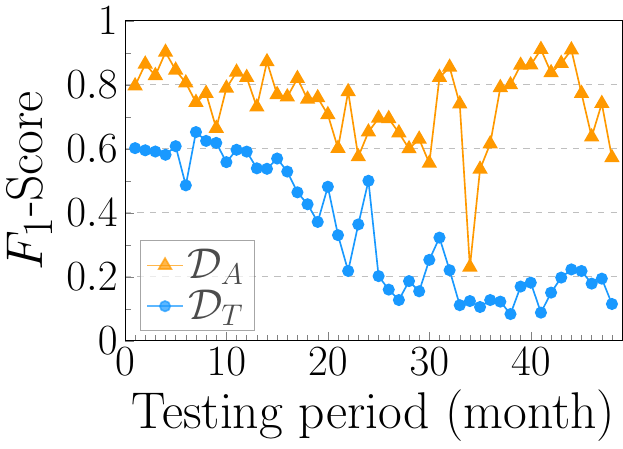}
        \subcaption{\ramda~\cite{ramda}}
    \end{minipage}
    \hfill
    \begin{minipage}{0.19\linewidth}
        \centering
        \includegraphics[width=\linewidth]{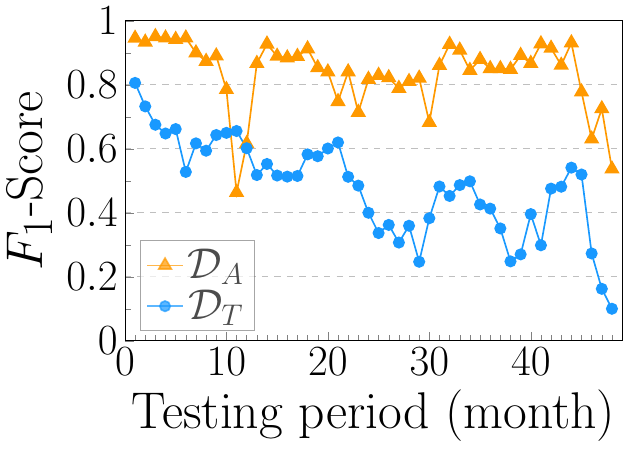}
        \subcaption{\HCC~\cite{chen2023continuous}}
    \end{minipage}
    \caption{\textbf{Motivational example.} Performance of five state-of-the-art classifiers on the two datasets from the same time frame, with the \fone{}s on \apigraph~\cite{apigraph_paper} consistently higher than the ones on \transcend~\cite{barbero2022transcending}. This paper investigates dataset bias factors and flawed evaluation strategies that may cause this discrepancy.}
    \label{fig:motivational_plot}
\end{figure*}

Let us consider two state-of-the-art Android malware datasets adopted in top-tier security venues: \textit{APIGraph}~\cite{apigraph_paper} and \textit{Transcendent}~\cite{barbero2022transcending}. Both were explicitly constructed to adhere to the spatio-temporal constraints proposed by \tesseract~\cite{pendlebury2019tesseract} and reflect recent trustworthy AI guidelines for dataset design and fair evaluations~\cite{dos_and_donts,flood2024bad}. Despite this, evaluation across five state-of-the-art malware detectors reveals striking discrepancies: \autoref{fig:motivational_plot} shows that detection models consistently achieve higher \fone{}s on \textit{APIGraph} than on \textit{Transcendent}, even when evaluated on the same time frame (2014--2018). From an \ac{ML} perspective, this indicates that sources of dataset bias persist beyond temporal and class-ratio considerations, and that hidden shifts or artifacts may systematically affect evaluation outcomes. These observations highlight the need for a deeper investigation into the nuanced factors that must be addressed to curate truly trustworthy datasets in cyber security domains.

More specifically, we identify five new spatio-temporal factors that can affect dataset sampling and evaluation of a learning-based Android malware detector. After introducing our hypotheses, we empirically evaluate the impact of each factor using five representative state-of-the-art Android malware classifiers: \drebin~\cite{arp2014drebin}, \deepdrebin~\cite{deepdrebin}, \malscan~\cite{malscan_paper}, \ramda~\cite{ramda}, and \HCC~\cite{chen2023continuous}. For each factor, we provide practical recommendations to reduce spatial and temporal biases. Finally, we survey existing Android malware datasets used in the literature to evidence how the factors we identify have often been ignored or overlooked in previous work. In fact, we discover that 95\% of the datasets we surveyed violate 3 or more of our recommendations, which can lead to biased experimental results.

\clearpage
In summary, we make the following contributions.

\begin{itemize}[noitemsep]
    \item We discover five novel spatio-temporal bias factors (\autoref{sec:preliminary-observations}) beyond \tesseract~\cite{pendlebury2019tesseract} and existing literature~\cite{dos_and_donts,sommer2010outside,flood2024bad}, which can lead to misleading results even if following all current recommendations for realistic evaluations, as shown in \autoref{fig:motivational_plot}. 
    
    \item We systematically analyze the impact of these factors on five state-of-the-art Android classifiers and two Android malware datasets that reflect current best practices. Where necessary, we augment our analysis by sampling additional data from \androzoo to further generalize our findings beyond the \textit{Transcendent} and \textit{APIGraph} datasets.

    \item We present actionable recommendations to address each identified factor, including an evaluation metric (\autoref{sec:factor-temporal-luck}) and sampling strategy (\autoref{sec:factor-sampling-size}) for evaluating and curating trustworthy Android malware datasets. 
    
    \item We conduct a prevalence study of the identified bias factors for popular Android malware datasets, highlighting that these are often overlooked (\autoref{sec:prevalence}).

\end{itemize}

To ensure reproducibility and foster bias-free evaluations in Android malware classification, we release our code and data publicly at \url{https://github.com/s2labres/hypercube-ml}.

\section{Preliminary Observations}\label{sec:preliminary-observations}
In this section, we investigate the possible causes behind the striking performance difference observed in~\autoref{fig:motivational_plot}. We reinforce that this occurs regardless of the feature representation or learning algorithm used and that the two datasets, \textit{APIGraph}~\cite{apigraph_paper} (\apigraph) and \textit{Transcendent}~\cite{barbero2022transcending} (\transcend), adhere to all spatial and temporal constraints outlined in \tesseract~\cite{pendlebury2019tesseract} and align with the recommendations by Arp et al.~\cite{dos_and_donts}. Moreover, they overlap in 2014--2018; hence, we restrict \apigraph to this period to match the time frame of \transcend, which allows for a direct comparison between the two datasets.

\paragraph{Family Overlap} Prior work~\cite{chow2023drift} showed that the appearance of new malware families is a major cause of drift in \transcend. Hence, we investigate the distribution differences between the two datasets by computing the ``family overlap,''~\ie the percentage of malware samples in the test set (from 2015 to 2018) belonging to families that existed in the training set (2014). A more formal definition of this metric can be found in Appendix~\autoref{sec:appendix:family-overlap-formula}. \autoref{fig:apigraph_transcendent_family_overlap} reports the family overlap for the two datasets, showing that new malware families steadily replace older ones in \transcend, while no clear trend can be observed in \apigraph. Considering that both datasets were sampled within the same time frame, we postulate that certain factors differed during the curation of the datasets, resulting in different distributions of malware samples.

\begin{figure}[t]
    \centering
    \includegraphics[scale=0.25]{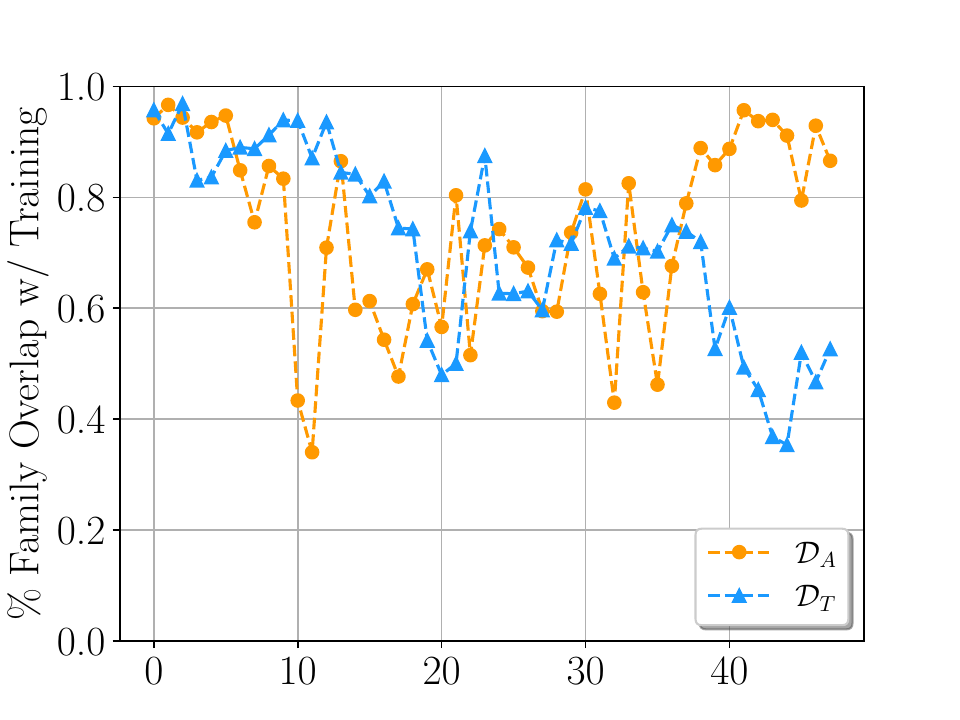}
    \caption{Monthly family overlap of \apigraph and \transcend at test time (2015--2018) with respect to families in the training set (2014).}
    \label{fig:apigraph_transcendent_family_overlap}
\end{figure}

\paragraph{Novel Spatio-Temporal Bias Factors} To identify possible sources of bias and explain the performance difference in \autoref{fig:motivational_plot}, we examine the high-level characteristics of \apigraph and \transcend in~\autoref{tab:datasets_differences}. We notice three major differences (Timestamp Types, VT Thresholds, and App Markets) and postulate that these may have contributed to the observed difference in performance. 

Specifically, \textit{App Markets} distribution affects the sampling source of both benign and malicious \acp{apk} (\autoref{sec:factor-app-markets}), while a higher \acf{vtt} may lead to malware that is ``easier to detect'' (\autoref{sec:factor-vtt}). Similarly, \textit{Timestamp Types} can influence both sampling and evaluation by inducing a dissimilar temporal distribution of \acp{apk} (\autoref{sec:factor-timestamps}).

Although the \textit{Dataset Size} is similar between \apigraph and \transcend, neither work clearly motivates the rationale behind its choice; in fact, none of the datasets we surveyed in our prevalence analysis (\autoref{sec:prevalence}) provide an explicit justification. Therefore, we study how dataset size may influence the statistical representativeness of a dataset (\autoref{sec:factor-sampling-size}). 

Finally, despite sharing the same \textit{Time Frame}, we argue that existing metrics such as Area Under Time (AUT)~\cite{pendlebury2019tesseract} capture only a temporal snapshot of performance. \daniel{TODO:Check} Particularly, it is possible that arbitrarily dividing a dataset into one training and one test sets may lead to a ``lucky'' split, resulting in a misleading performance for a given time frame---a phenomenon we call \textit{Temporal Luck} (\autoref{sec:factor-temporal-luck}).

In the remainder of this paper, we analyze each factor by formulating hypotheses and evaluating their impact on the resulting data distribution and on the detection performance of the classifiers we consider. From our findings, we determine whether each factor introduced bias during the curation and evaluation of \apigraph and \transcend and, then, provide practical recommendations on how to prevent this. We further quantify the prevalence of each factor in the scientific literature by surveying Android malware datasets in previous work and assessing them with respect to our recommendations (\autoref{sec:prevalence}).

\begin{table}[t]
\centering
\scriptsize
\caption{High-level differences between \apigraph and \transcend.}
\label{tab:datasets_differences}
\scalebox{0.85}{
\begin{tabular}{lp{3cm}p{3cm}}
\textbf{Characteristic} & \apigraph~\cite{apigraph_paper} & \transcend~\cite{barbero2022transcending} \\ 
\toprule
Time Frame & 2012-2018 & 2014-2018 \\ \midrule
Timestamp & \vtdates & AZ \dexdates \\ \midrule
\shortvt Threshold & 15 & 4 \\ \midrule
\multirow{3}{*}{App Markets} & \googleplay (goodware), & \googleplay~(91\%), \\
                                                  & VirusShare (malware) & Anzhi~(7\%), \\
                                                & & AppChina~(2\%) \\ \midrule
Dataset Size & 320,001 & 259,230 \\  \midrule
Dataset Size (2014-18) & 241,611 & 259,230 \\ \midrule
Num. Families & 484 & 492 \\ \bottomrule
\end{tabular}}
\end{table}

\section{Experimental Settings} \label{sec:experimental-setup}
Before introducing and analyzing the impact of the factors we identify, we provide a brief overview of the datasets, classifiers, and metrics used in the remainder of the paper.

\paragraph{Datasets}
To investigate the five identified factors, we employ \apigraph~\cite{apigraph_paper} and \transcend~\cite{barbero2022transcending} (see~\autoref{sec:preliminary-observations}). Where relevant, we generalize our findings using metadata from \androzoo~\cite{androzoourl}, reports from \longvt~\cite{vturl}, and family labels unified with Euphony~\cite{hurier2017euphony}. We sample across multiple markets, recreate both \apigraph and \transcend with different \acp{vtt} at different time points, and curate datasets with varying strategies to determine appropriate sample sizes. Further details are provided in \autoref{sec:factor-app-markets}, \autoref{sec:factor-vtt}, \autoref{sec:factor-sampling-size}.

\paragraph{Classifiers}
All our experiments are carried out on five representative Android malware classifiers:

\begin{itemize}[noitemsep, topsep=0pt, parsep=0pt]
\item \textbf{\drebin}\cite{arp2014drebin}: Linear SVM using binary features from static analysis (\eg APIs, URLs, Activities).
\item \textbf{\deepdrebin}\cite{deepdrebin}: MLP using the \drebin features.
\item \textbf{\malscan}\cite{malscan_paper}: Graph-based approach using degree centrality with a Random Forest classifier.
\item \textbf{\ramda}\cite{ramda}: DNN combining a VAE and an MLP, focusing on sensitive APIs, Permissions, and Intents.
\item \textbf{\HCC}~\cite{chen2023continuous}: Encoder-guided embedding using malware family labels for enhanced class separability.
\end{itemize}
Appendix~\ref{sec:appendix:classifier-details} provides more details on these classifiers. Notice that their diversity allows for robust and representation-agnostic insights. Since \HCC shows similar trends yet consistently outperforms the other classifiers, we show performance plots only for \HCC throughout the remainder of the paper but include all the other performance plots in the Appendix. 

\paragraph{Metrics}
We assess each factor's impact on classification performance using the standard performance metrics \fone and \textit{AUT} for temporal evaluations~\cite{pendlebury2019tesseract}. Additionally, we report the \textit{Family Overlap} measure introduced in~\autoref{sec:preliminary-observations} to quantify the distribution shift of malware families independent of the feature representation or classifier used (see Appendix~\autoref{sec:appendix:family-overlap-formula}).

\section{Factor 1: Timestamp Types} \label{sec:factor-timestamps}

\subsection{Hypothesis} Since \tesseract advocated for time-aware evaluations~\cite{pendlebury2019tesseract,allix2015your,miller2016reviewer}, the integration of temporal information has become an essential consideration in constructing realistic datasets. In particular, we note that timestamps do not just impact the temporal order of samples but the sampling process itself. For example, in \apigraph, samples with VirusTotal First-Submission Dates (\vtdates) before 2012 and after 2018 were discarded; and in \transcend, the \androzoo population was filtered by excluding samples with \dexdate newer than 2018 or older than 2014. 
As such, timestamp-based filtering directly shapes the resulting data population; thus, we analyze the impact of commonly used timestamp types on the construction of time-based datasets.

For the Android domain, several timestamps are available to practitioners. Among the commonly-used ones, \vtdates (used in \apigraph) indicate when \acp{apk} were first submitted to VirusTotal~\cite{vturl}, while \dexdates correspond to the last modification date of an \ac{apk}'s \textit{classes.dex} file (the executable code of the Android app). Other relevant timestamps include \googleplay \uploaddates, namely the date in which an \ac{apk} was uploaded to \googleplay, and \androzoo \crawldates, which indicate the date on which \androzoo crawled an \ac{apk} from a market.

All these timestamps provide different temporal information and can be loosely sorted into three categories: \textit{Creation Timestamps}, \textit{Publication Timestamps}, and \textit{Third-party Timestamps}. Creation Timestamps aim to capture the date an \ac{apk} was created or built. Among the notable timestamps, \dexdates are the only one falling in this category. \dexdates are known to be unreliable~\cite{li2018moonlightbox}, as they are easily modifiable by attackers and removed by default by all Android SDKs released after 2016~\cite{dex_dates_2016}. Publication timestamps capture the date on which an \ac{apk} was uploaded to a public market such as \googleplay. Since they are market-specific, \acp{apk} uploaded to multiple markets will have multiple publication timestamps. Finally, Third-party Timestamps capture the date an \ac{apk} was included in an app repository managed by professionals/academics, such as \androzoo and \acl{vt}.

Given that timestamps provide different temporal information, it is important to understand the distributions they model. Creation Timestamps reflect the evolution of malware families and attack techniques over time but may misrepresent reality since republished malware would be assigned to an earlier point despite reappearing in the present. We obtained confirmation from industrial partners that they do see old malware still circulating years after its original appearance. Publication Timestamps instead capture the population of an app market at a specific moment, closely matching the deployment setting of antivirus engines. Third-party Timestamps are less reliable indicators of real-world distributions, as samples can be uploaded before or after market release. Nevertheless, they may be useful proxies when other timestamps are unavailable. 

In this section, we first evaluate the ability of Creation Timestamps to model the evolution of malware, knowing that old malware may appear at a later time.

\daniel{TODO: Shall we use different names of H1 and H2?--- fair point... Any suggestions? Mario}
\hypothesisbox{\textbf{Hypothesis 1: \factorTimestampName}}{Due to the reappearance of old malware, using Creation Timestamps to sample time-aware datasets within time frame $[T_1, T_2]$ ($T_1 < T_2$) may yield different malware distributions, depending on \textit{when} sampling is conducted.}

\noindent The intuition behind this hypothesis is that as old malware reappears, the underlying population may change from the one observed in the past. 

We then evaluate the ability of Third-party Timestamps to emulate Publication Timestamps. We would expect \acp{apk} to enter third-party services such as \androzoo and \ac{vt} after their publication on an official app market. In particular, we investigate whether the temporal distribution of a third-party service and a market may be similar but misaligned, which would still allow the Third-party Timestamps to fairly represent that market's distribution. 

\hypothesisbox{\textbf{Hypothesis 2: \factorTimestampName}}{Third-Party Timestamps may fail to accurately reflect the temporal distribution of an app market.}

\begin{figure}[t]
  \centering
  \includegraphics[scale=0.35]{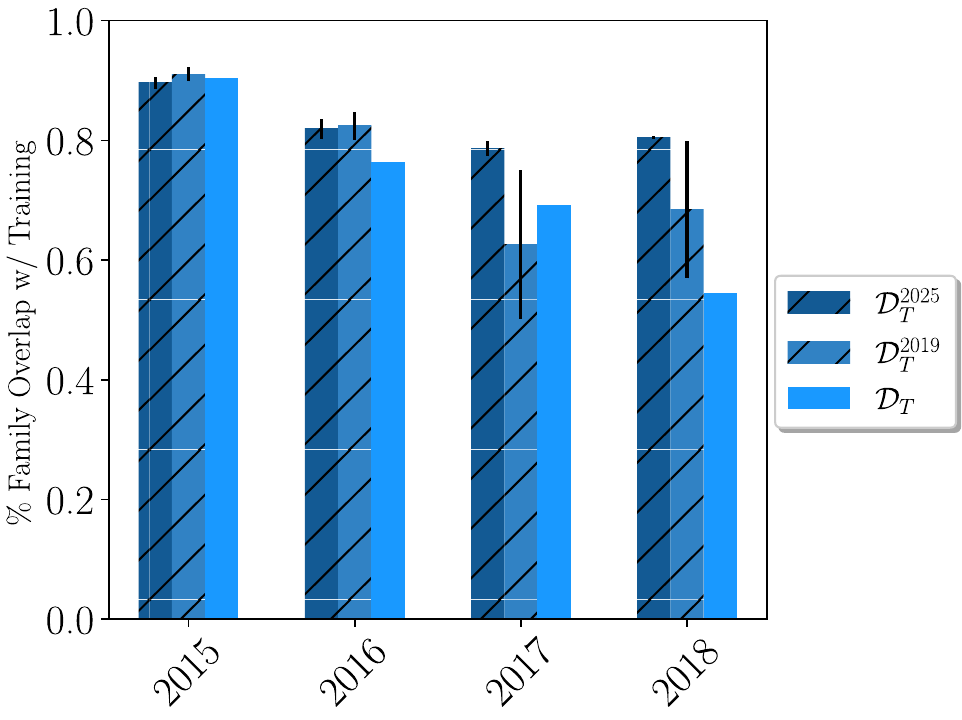}
  \caption{Yearly family overlap of 3 datasets identically and independently sampled from a snapshot of \androzoo in 2025 and in 2019, using the sampling parameters of \transcend.}\label{fig:transcend_2025}
\end{figure}

\subsection{Impact} \label{sec:timestamps-impact}

\paragraph{Hypothesis 1}
To validate this hypothesis, we first resample \transcend in May 2025 three times. We then leverage \androzoo \crawldates to reconstruct a snapshot of \androzoo when \transcend was sampled in 2019, by filtering out samples that were not present on \androzoo at that time. For both configurations, we independently sample three datasets, maintaining the same number of malware per month as \transcend, using \dexdates for sampling and ordering, and \ac{vtt}=4. We analyze the resulting malware distributions for the two configurations by averaging the family overlap of each. Further details regarding how we recreated \transcend can be found in Appendix~\autoref{sec:appendix:datasets}. 

\autoref{fig:transcend_2025} reports the family overlap for different samplings. It can be observed that the datasets sampled from \androzoo in May 2025 ($D_{T}^{2025}$) show a higher family overlap in 2017 and 2018, with significantly smaller standard deviation. In contrast, datasets sampled from a snapshot of \androzoo when \transcend was originally sampled ($D_{T}^{2019}$) present a lower average family overlap in the last two years, with a higher standard deviation, indicating greater variability.

These findings confirm our first hypothesis: sampling a dataset from a fixed time frame through Creation Timestamps will yield different malware distributions depending on \textit{when} the sampling is performed (in this case, 2019 vs. 2025). Moreover, this further highlights that old malware may reappear in a population at a later time than its original creation, making the task of modeling the evolution of malware variants through simple Creation Timestamps inaccurate.

\paragraph{Hypothesis 2} To validate our second hypothesis, we compare the temporal distribution of \googleplay Publication Timestamps (\uploaddates) with Third-party Timestamps from \androzoo (\crawldates) and \acl{vt} (\vtdates). We model this distribution as a binary matrix of shape $(N_T,|D_T|)$, where $N_T$ is the number of time units within the time frame $T=[T_1, T_2]$ (\eg 12 months from 1-1-2021 to 31-12-2021), and $|D_T|$ is the number of samples published on \googleplay in $[T_1, T_2]$. Each entry $\left(i, j\right)$ equals $1$ if sample $j$ was published on the market or uploaded to a third-party service during the $i^{th}$ time slot, $0$ otherwise. Each column in the matrix represents an individual \ac{apk} and can sum to $0$ or $1$. For a third-party service, a column can sum to $0$ if the corresponding \ac{apk} was uploaded outside of $T$.

We reuse the metadata of 430k \acp{apk} that we collected for the experiments in~\autoref{sec:appendix:dataset-size}. These \ac{apk} were published on \googleplay between 1-1-2021 and 31-12-2023 and are available on \androzoo. Using \uploaddates, \vtdates, and \crawldates, we construct yearly temporal distributions for \googleplay, \ac{vt}, and \androzoo, respectively. To assess the similarity of the three temporal distributions, we compute the \textit{cosine similarity} between the \googleplay distribution and those of \ac{vt} and \androzoo. \autoref{tab:cosine_similarity_by_year}~reports the monthly cosine similarities with their standard deviations. The results show that the monthly mean values of \androzoo \crawldates are closer to $1$ in 2021 and 2023, while those of \acl{vt} \vtdates are closer to $0$ for all three years, indicating dissimilarity from \googleplay \uploaddates.

\begin{table}[t]
\caption{\textbf{Monthly mean cosine similarity between \googleplay \uploaddates and two timestamp distributions, \ie \acl{vt} \vtdates and \androzoo \crawldates}. 1 indicates identical, while -1 indicates inversely correlated vectors. \crawldates are closer to 1 compared to \vtdates, suggesting \vtdates are more dissimilar to \uploaddates.}
\scriptsize
\label{tab:cosine_similarity_by_year}
\centering
    \begin{tabular}{l|rr|rr}
    & \multicolumn{2}{c}{\textbf{Goodware}} & \multicolumn{2}{c}{\textbf{Malware}} \\
    & \vtdates & \crawldates & \vtdates & \crawldates \\
    \hline
    2021 & 0.20$\pm$0.09 & 0.76$\pm$0.04 & 0.23$\pm$0.10 & 0.71$\pm$0.06 \\
    2022 & 0.13$\pm$0.07 & 0.24$\pm$0.05 & 0.14$\pm$0.07 & 0.67$\pm$0.08 \\
    2023 & 0.18$\pm$0.06 & 0.63$\pm$0.13 & 0.21$\pm$0.08 & 0.58$\pm$0.13 \\ \bottomrule
    \end{tabular}
\end{table}

However, the temporal distributions of \ac{vt} and \androzoo may be misaligned with respect to \googleplay, as \acp{apk} are expected to enter these services only after publication on \googleplay. To verify this, we align each pair of temporal distributions by solving the \ac{dtw} problem~\cite{time_series_analysis} and report the final cumulative distance (normalized as in~\cite{dtw}) in~\autoref{tab:dtw_by_year}. For reference, we also compute the \ac{dtw} distance between the \googleplay distribution and a random one, constructed by assigning each \ac{apk} a random timestamp succeeding its original \uploaddate. Both \ac{vt} and \androzoo distances are close to those of a random distribution. Notably, the \ac{vt} distance for malware is identical to that of a random distribution in 2023 and nearly identical in 2022. These results indicate that Third-party Timestamps cannot reliably represent the temporal distribution of \googleplay.

\begin{table}[t]
\caption{\textbf{\ac{dtw} distance between \googleplay \uploaddates and three timestamp distributions, \ie \acl{vt} \vtdates, \androzoo \crawldates, and random temporal distributions.} Values in brackets are normalized against the random distribution, with lower values indicating that the time-series aligns better with \uploaddates. }
\label{tab:dtw_by_year}
\centering
\scalebox{0.7}{
   \begin{tabular}{l|rrr|rrr}
   & \multicolumn{3}{c}{\textbf{Goodware}} & \multicolumn{3}{c}{\textbf{Malware}} \\
   & Random & \vtdates & \crawldates & Random & \vtdates & \crawldates \\
   \hline
   2021 & 116 (1.00)& 94 (0.81) & 71 (0.61)&  52 (1.00) & 44 (0.85) & 34 (0.65) \\
   2022 & 111 (1.00)& 93 (0.84) & 68 (0.61) & 46 (1.00) & 45 (0.98) & 35 (0.76) \\
   2023 & 100 (1.00)& 90 (0.90) & 78 (0.78) & 38 (1.00) & 38 (1.00) & 34 (0.90) \\ \bottomrule

   \end{tabular}}
\end{table}

\subsection{Recommendations}
\recommendationboxnew{Recommendation: \factorTimestampName}{Publication Timestamps \textbf{must} be used to model the population of a market (\eg \googleplay \uploaddates).}

\section{Factor 2: Temporal Luck}\label{sec:factor-temporal-luck}
\subsection{Hypothesis}
Two datasets, $D_1$ and $D_2$, sampled from different time frames (\eg 2012--2014 and 2016--2018), can be expected to represent different distributions (\eg they may vary in terms of family composition). Therefore, no meaningful conclusions can be drawn from comparing the performance of a malware classifier on $D_1$ with the performance of another malware detection method evaluated on $D_2$. This issue of comparing malware detection methods on different datasets presenting a distribution mismatch may arise when researchers directly compare their results with those reported in other related papers (``comparison on paper'').

Given two malware detection approaches $A_1$ and $A_2$, a straightforward solution would be to evaluate both on the \textit{same} dataset, as done in recent work (\eg ~\cite{apigraph_paper,malscan_paper,barbero2022transcending}). Nevertheless, this may still not suffice to conclude which method performs best, as the reported results may still be influenced by what we call \textit{\factorTemporalLuckName}. This phenomenon arises when an arbitrary time-aware split into training and testing data inflates the performance of a given method, due to the training set being ``exceptionally'' good or the subsequent testing period not drifting significantly. In practical terms, given a dataset $\mathcal{D}_{2012-2014}$, $A_1$ may perform better than $A_2$ when training on 2012 and testing on 2013, whereas $A_2$ may exhibit superior performance when training on 2013 and testing on 2014. Hence, we postulate the following hypothesis.

\hypothesisbox{\textbf{Hypothesis: \factorTemporalLuckName}}{Given a dataset, the performance of a malware detection method may vary based on the temporal train/test splits.}

\begin{figure}[t]
    \begin{minipage}{0.49\linewidth}
        \centering
        \includegraphics[width=\textwidth]{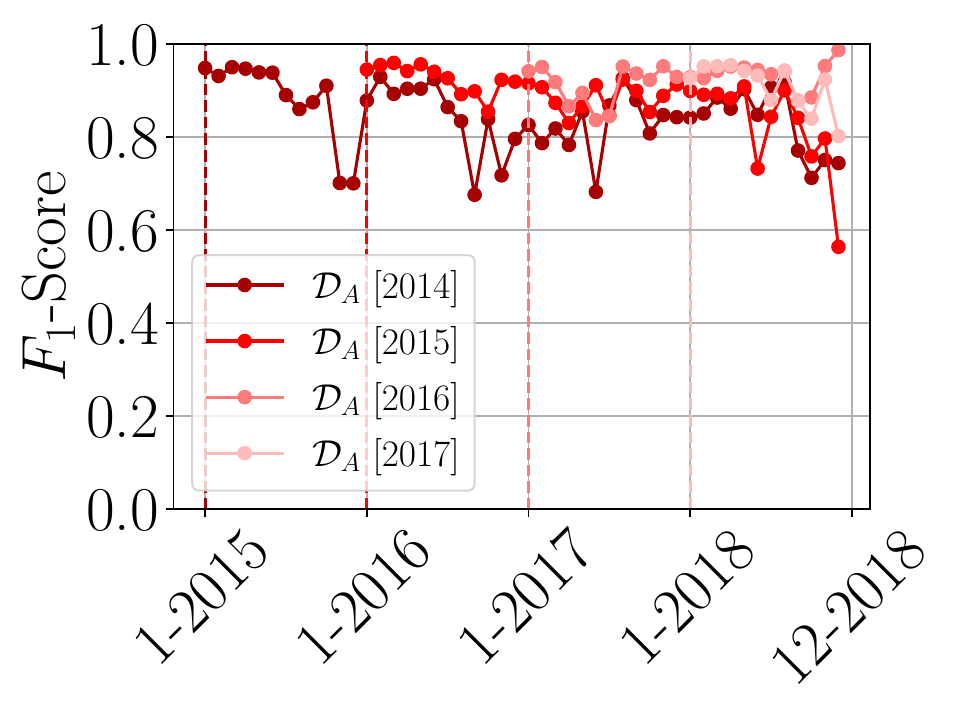}
        \subcaption{\apigraph}    
    \end{minipage}
    \hfill
    \begin{minipage}{0.49\linewidth}
        \centering
        \includegraphics[width=\linewidth]{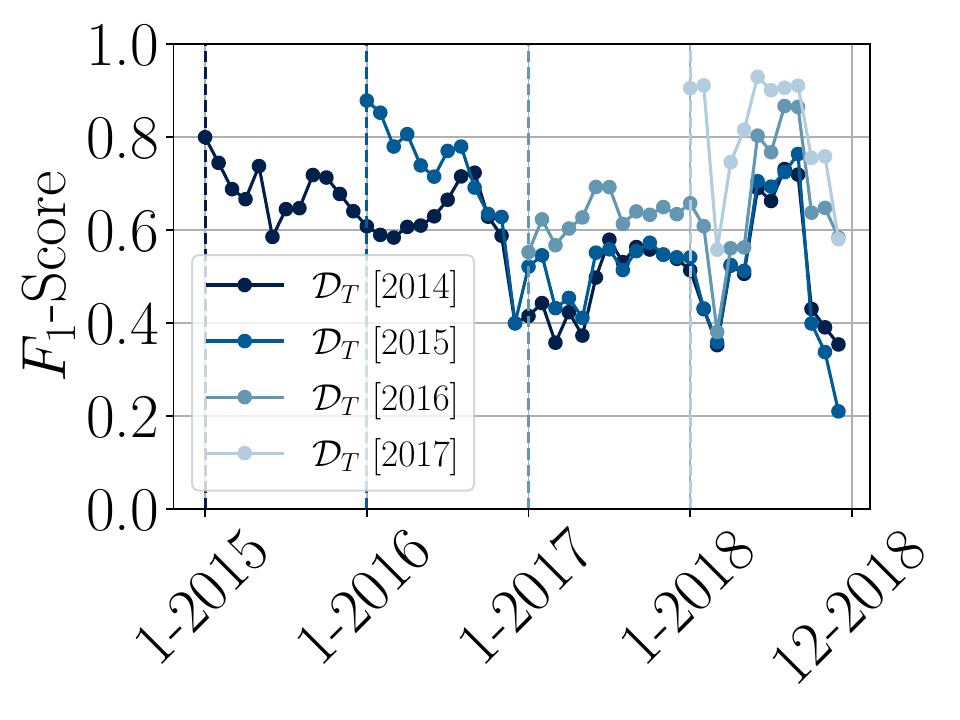}
        \subcaption{\transcend}
    \end{minipage}
    \caption{\textbf{Impact of \factorTemporalLuckName on the performance of \HCC.}  Each line indicates the \fone of a model trained on each year between 2014 and 2017. It can be seen that training on different years within the same dataset can yield different performance profiles. The issue is evident on \transcend, but also present in \apigraph. For other classifiers, refer to Appendix~\autoref{sec:appendix:tempora_luck_example}.}
    \label{fig:temporal_luck_motivation_hcc}
\end{figure}

\begin{table}[t]
\caption{\textbf{\factorTemporalLuckName.} 12-month window \ac{AUT}: 2014-2015 indicates that the model is trained on 2014 and its \ac{AUT} value computed on 2015. The performance of a model on a given dataset changes depending on how we temporally split the dataset.}
\label{tab:aut_by_year}
\scriptsize
\centering
    \begin{subtable}{\linewidth}
        \begin{tabular}{lrrrr|rr}
        
        Classifier & \makecell{\rotatebox{60}{2014-2015}} & \makecell{\rotatebox{60}{2015-2016}} & \makecell{\rotatebox{60}{2016-2017}} & \makecell{\rotatebox{60}{2017-2018}} & $\mu_{AUT}$ & $\sigma_{AUT}$ \\ \toprule
        \drebin & 0.85 & 0.91 & 0.90 & 0.87 & 0.88 & 0.02 \\ 
        \deepdrebin & 0.84 & 0.93 & 0.91 & 0.91 & 0.90 & 0.03\\ 
        \malscan & 0.67 & 0.65 & 0.66 & 0.87 & 0.72 & 0.09\\ 
        \ramda & 0.75 & 0.81 & 0.77 & 0.59 & 0.73 & 0.09 \\
        \HCC & \textbf{0.89} & \textbf{0.93} & \textbf{0.91} & \textbf{0.92}  & \textbf{0.91} & \textbf{0.01}\\ \bottomrule
        \end{tabular}
    \subcaption{\apigraph}
    \end{subtable}
\vspace{4pt}
    \begin{subtable}{\linewidth}
        \begin{tabular}{lrrrr|rr}
        Classifier & \makecell{\rotatebox{60}{2014-2015}} & \makecell{\rotatebox{60}{2015-2016}} & \makecell{\rotatebox{60}{2016-2017}} & \makecell{\rotatebox{60}{2017-2018}} & $\mu_{AUT}$ & $\sigma_{AUT}$ \\ \toprule
        \drebin & \textbf{0.71} & \textbf{0.75} & \textbf{0.63} & 0.80 & \textbf{0.72} & \textbf{0.06}  \\ 
        \deepdrebin & 0.69 & 0.74 & 0.55 & 0.82 & 0.70 & 0.10\\ 
        \malscan & 0.45 & 0.42 & 0.53 & \textbf{0.83} & 0.56 & 0.16\\
        \ramda & 0.54 & 0.56 & 0.44 & 0.74 & 0.57 & 0.11\\
        \HCC & 0.70 & \textbf{0.75} & \textbf{0.63} & 0.80 & \textbf{0.72} & \textbf{0.06}\\ \bottomrule
        \end{tabular}
    \subcaption{\transcend}
    \end{subtable}
    \vspace{-6pt}
\end{table}

\subsection{Impact}
\autoref{fig:temporal_luck_motivation_hcc} provides an overview of the impact of \factorTemporalLuckName on both \apigraph and \transcend. We employ the following settings: we train classifiers on data from each year in 2014--2018 and test them on the following years. In practice, we divide the datasets into four subsets: the first having 2014 as training window and testing on 2015--2018, the second having 2015 for training and 2016--2018 for testing, and so on. It is visually evident in \transcend that significantly different performance profiles may be obtained depending on the temporal split. This can also be observed for \apigraph, albeit to a lesser extent.

To quantify \factorTemporalLuckName in \apigraph and \transcend, we compute the \ac{AUT} when training on one year and testing only on the following one (\ie (i) training on 2014 and testing on 2015, (ii) training on 2015 and testing on 2016, etc.). The \ac{AUT} summarizes the performance of a classifier over a certain time frame, and in the presence of \factorTemporalLuckName, it can emphasize when a proposed approach works consistently across different temporal splits. The results summarized in \autoref{tab:aut_by_year} show again significantly different results on \transcend for all classifiers and temporal configurations, and smaller differences for \apigraph.

The \ac{AUT} can reveal substantial performance variation across training years. For example, all classifiers show a significantly lower performance in 2016--2017 than in 2017--2018 of \transcend. However, the \ac{AUT} may complicate the comparison between classifiers, as no individual classifier may perform better than the others in all scenarios (with the exception of \HCC for \apigraph). For example, \malscan performs better than both \drebin and \HCC in 2017--2018 of \transcend. However, the reverse is true for the other three temporal configurations. Similarly, \deepdrebin performs better than \drebin in 2015--2016, 2016--2017, and 2017--2018, but worse in 2014--2015 of \apigraph.

To facilitate direct comparison between different malware detection methods, we propose the \ac{A-AUT}, an extension of the \ac{AUT} metric that provides a compact summary of the time-aware performance of a given classifier (given a metric $f$, such as the \fone), while accounting for \factorTemporalLuckName:

\begin{equation}
    \text{A-AUT}_f(\mathcal{C},T, E) = \bigl(\mu_{{AUT}_f}, \sigma_{{AUT}_f}\bigr),
\label{eq:avg_aut}
\end{equation}
where $\mu_{{AUT}_f}$ and $\sigma_{{AUT}_f}$ are the mean and standard deviation of $\text{AUT}_f(C_{t_i}, e_i)$, computed by training classifier $\mathcal{C}$ on each $t_i \in T$ and evaluating on the corresponding $e_i \in E$, with:
\begin{align}
    T &:= \{D_{in_T:(i+1)n_T-1}\}_{i=0}^{k-1} \\
    E &:= \{D_{(i+1)n_T:(i+1)n_T+n_E-1}\}_{i=0}^{k-1} \\
    k &= \lfloor(|D| - n_E) / n_T\rfloor.
\end{align}

More specifically, a dataset $\mathcal{D}$ is divided into $k$ temporally successive training sets ($T$) as well as into $k$ evaluation sets ($E$), having sizes $n_T$ and $n_E$, respectively, with $n_E \geq n_T$. The formula for the case $n_T > n_E$ is described in Appendix~\autoref{sec:appendix:tempora_luck_example}. For each dataset pair $(t_i,e_i)\in \{(t_0,e_0), \cdots, (t_{|D|/n_T}, e_{|D|/n_T})\}$, with $t_i\in T$ and $e_i\in E$, we train the classifier $\mathcal{C}$ on $t_i$ and compute its AUT on $e_i$. Notice that each $e_i$ starts at time unit $in_T$ and includes the following $n_E$ time units. A practical example of this \textit{rolling window} approach is provided in \autoref{sec:appendix:tempora_luck_example}.

In \autoref{tab:aut_by_year}, we report the A-AUT of the \fone computed for our considered classifiers and datasets, setting both $n_T$ and $n_E$ to one year. The proposed metric facilitates the direct comparison of different malware detection methods while accounting for Temporal Luck. For example, it can help determine which classifier, \drebin or \deepdrebin, performs better on average on \transcend: \drebin performs slightly better than \deepdrebin ($\mu_{AUT}=0.72$ versus $\mu_{AUT}=0.70$) while also exhibiting more stable performance across different temporal splits ($\sigma_{AUT}=0.06$ versus $\sigma_{AUT}=0.10$). \malscan shows weaker performance than both, with higher variability. \HCC achieves the highest $\mu_{AUT}$ and lowest $\sigma_{AUT}$ for \apigraph. For \transcend, \HCC and \drebin share the best performance with $\mu_{AUT}=0.72$ and $\sigma_{AUT}=0.06$.

\subsection{Recommendation}
As a single train-test split may unfairly (dis)advantage one classifier over another, we recommend using the proposed \ac{A-AUT} metric to provide a more robust mean estimate of the classifiers' performance, along with a measure of performance stability over time splits.

\recommendationboxnew{Recommendation: \factorTemporalLuckName}{\ac{A-AUT} \textbf{must} be used to compare malware detection methods.
If multiple methods have comparable \acp{A-AUT}, an AUT breakdown \textbf{must} be provided. \ac{A-AUT} \textbf{may} be used with 12-month windows to obtain a yearly summary of the results.}

\section{Factor 3: App Markets}\label{sec:factor-app-markets}
\subsection{Hypothesis}

Android apps are distributed across multiple markets. For example, \googleplay is the dominant platform in western regions, while other markets are more prevalent in eastern countries.
When constructing Android malware datasets, recent studies~\cite{pendlebury2019tesseract,barbero2022transcending,apigraph_paper,chen2023continuous} adopt multi-market sampling to increase data diversity and scale. However, this approach may introduce artifacts during training, causing classifiers to learn market-related features instead of security-relevant ones~\cite{dos_and_donts}.

\autoref{tab:markets_metadata_azoo_datasets} shows the market composition of \androzoo, \apigraph, and \transcend, based on metadata from \androzoo~\cite{androzoourl}. The data reveal a substantial imbalance: certain markets contribute disproportionately to the \androzoo population, and \apigraph and \transcend differ distinctly in their sampling. In particular, \apigraph exhibits a pronounced sampling bias, sourcing malware primarily from VirusShare and goodware from \googleplay.

Therefore, we hypothesize that an imbalanced market composition in the training data, characterized by malware and goodware being predominantly sampled from different markets, causes statistically significant performance degradation when evaluated on market-balanced or differently composed test sets.

\hypothesisbox{Hypothesis: \textbf{\factorMarketsName}}{Altering the composition of \factorMarketsName used in dataset sampling significantly affects goodware and malware class-distribution and leads to reduced detection performance (\eg \fone or AUT) when models are evaluated on out-of-sample market distributions.}

\begin{table}[t]
\footnotesize
    \centering
    \caption{\textbf{Comparison of market sources for the \acp{apk} within \androzoo, \apigraphbold and \transcendbold.} The market information of each \ac{apk} is taken from the \androzoo metadata (note that each \ac{apk} can be present in multiple markets).}
    \label{tab:markets_metadata_azoo_datasets}
    \scalebox{0.7}{
    \begin{tabular}{lrrrrrr}
        Market & \multicolumn{2}{c}{\androzoo} & \multicolumn{2}{c}{\textbf{\apigraph}} & \multicolumn{2}{c}{\textbf{\transcend}} \\
        & Goodware & Malware& Goodware & Malware & Goodware & Malware \\
        \toprule
        angeeks & 0.18\% & 0.64\% & 0.21\% & 0.34\% & - & - \\
        anzhi & 1.64\% & \textbf{29.25\%} & 0.37\% & 1.89\% & \textbf{3.24\%} & \textbf{37.57\%} \\
        apk\_bang & - & - & - & 0.01\% & - & - \\
        appchina & \textbf{2.57\%} & \textbf{22.74\% }& \textbf{1.75\%} & 3.75\% & 2.28\% & \textbf{11.28\%} \\
        fdroid & 0.27\% & 0.01\% & - & - & 0.08\% & - \\
        freewarelovers & 0.02\% & - & 0.03\% & - & - & - \\
        genome & - & 0.06\% & - & 0.57\% & - & - \\
        hiapk & 0.01\% & 0.04\% & 0.01\% & 0.01\% & 0.02\% & 0.06\% \\
        mi.com & 0.13\% & 2.82\% & 0.06\% & 0.23\% & 0.4\% & 4.12\% \\
        PlayDrone  & \textbf{5.75\%} & 6.41\% & \textbf{33.52\%} & 9.75\% & \textbf{13.26\%} & 6.35\% \\
        \googleplay & \textbf{94.91\%} & \textbf{29.50\% }& \textbf{99.99\%} & \textbf{10.13\%} &\textbf{ 95.33\%} & \textbf{49.29\%} \\
        praguard  & - & 0.47\% & - & 0.01\% & - & - \\
        proandroid & 0.02\% & 0.01\% & 0.06\% & 0.02\% & - & - \\
        slideme  & 0.21\% & 0.27\% & 0.3\% & 0.14\% & 0.02\% & - \\
        unknown  & 0.01\% & 1.27\% & 0.08\% & \textbf{23.74\%} & 0.02\% & 0.49\% \\
        VirusShare & 0.12\% & 15.95\% & 0.07\% &\textbf{ 81.93\%} & 0.14\% & 2.05\% \\
        1mobile  & 0.2\% & 0.48\% & 1.36\% & 0.35\% & - & - \\
        \bottomrule
    \end{tabular}}
\end{table}

\subsection{Impact}
\begin{figure}[t]
    \begin{minipage}{0.49\linewidth}
        \centering
        \includegraphics[width=\textwidth]{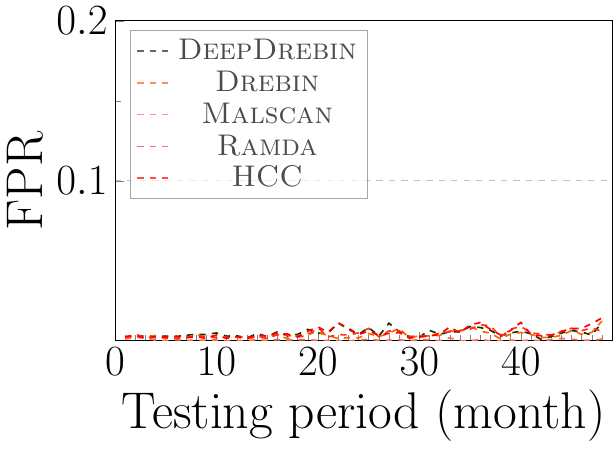}
        \subcaption{\apigraph}    
    \end{minipage}
    \hfill
    \begin{minipage}{0.49\linewidth}
        \centering
        \includegraphics[width=\linewidth]{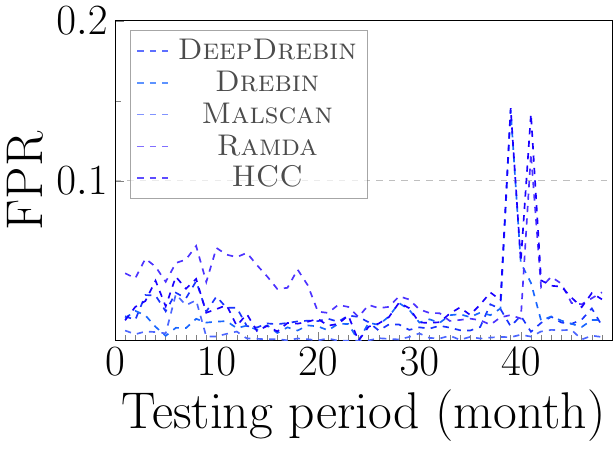}
        \subcaption{\transcend}
    \end{minipage}
    \caption{\ac{fpr} of the five classifiers trained on 2014 and evaluated across 2015--2018.}
    \label{fig:fprs}
\end{figure}

To investigate whether the performance on \apigraph is influenced by goodware being sampled entirely from \googleplay and malware primarily from VirusShare, we first analyze the \ac{fpr} of both datasets. As expected, classifiers trained on \apigraph tend to be steadily confident in predicting benign \acp{apk}. In contrast, classifiers trained on \transcend exhibit much more variable \acp{fpr}. \autoref{fig:fprs} shows the \ac{fpr} trends of all five classifiers when trained and evaluated on both datasets.

For a more systematic analysis, we now empirically evaluate the impact of market composition by constructing six dataset configurations, designed to simulate the influence of market sources on dataset composition, and evaluating their effect on malware detection performance. We sample all these datasets from \androzoo; in particular, from two groups of markets. We collect samples only from \googleplay (\textit{GP}), as it is the dominant market in \androzoo, and we group all remaining 3rd-Party Markets together (\textit{3PM}), as 3PM are otherwise individually too small (compared to GP). In total, we collected 15,000 goodware and 15,000 malware samples from each of \googleplay and 3PM, summing to 60,000 samples.
A detailed breakdown of these configurations is given in Appendix~\autoref{sec:appendix:datasets}. Each dataset configuration consists of 20,000 samples for training and 5,000 samples for testing, randomly selected from the collected pool of samples. Due to the limited number of samples per time split, we do not perform a temporal evaluation and instead sample three datasets per configuration and report the average \fone.

\begin{table}[t]

\caption{\textbf{Evaluation results of sampling from diverse app market sources.} Training and testing on the same markets consistently yields higher \fone compared to training and testing on different markets, showing \acp{apk} from multiple markets are inherently different. Mixing markets can lead to unclear evaluation results skewed towards the majority market. Combining malware and goodware from separate markets inflates performance, corroborating findings in~\cite{dos_and_donts}. Standard deviation for all results is lower than 0.03.}
    \label{tab:data_market_results}
\centering
\scalebox{0.7}{
    \begin{tabular}{l|ll|rrrrr} 
    
    &\multicolumn{2}{c}{}  & \textbf{\drebin}  &  \textbf{\deepdrebin} & \textbf{\malscan} & \textbf{\ramda} & \textbf{\HCC}\\
   No. & \textbf{Train Set} & \textbf{Test Set} & \fone & \fone & \fone & \fone & \fone\\ \toprule

1&$\mathcal{D}_{GP}$ & $\mathcal{D}_{GP}$&$0.776$&$0.788$&$0.841$&$0.524$&$0.813$\\
2&$\mathcal{D}_{3PM}$ & $\mathcal{D}_{3PM}$&$0.658$&$0.699$&$0.614$&$0.514$&$0.782$\\\\
3&$\mathcal{D}_{GP}$ & $\mathcal{D}_{3PM}$&$0.265$&$0.247$&$0.298$&$0.227$&$0.261$\\
4&$\mathcal{D}_{3PM}$ & $\mathcal{D}_{GP}$&$0.415$&$0.290$&$0.416$&$0.262$&$0.348$\\\\

5&$\mathcal{D}_{GP}$ & $\mathcal{D}_{EVEN}$&$0.422$&$0.395$&$0.511$&$0.344$&$0.417$\\
6&$\mathcal{D}_{3PM}$ & $\mathcal{D}_{EVEN}$&$0.622$&$0.558$&$0.581$&$0.454$&$0.642$\\\\

7&$\mathcal{D}_{EVEN}$ & $\mathcal{D}_{GP}$&$0.740$&$0.747$&$0.789$&$0.576$&$0.799$\\
8&$\mathcal{D}_{EVEN}$ & $\mathcal{D}_{3PM}$&$0.659$&$0.566$&$0.666$&$0.467$&$0.781$\\
9&$\mathcal{D}_{EVEN}$ & $\mathcal{D}_{EVEN}$&$0.659$&$0.706$&$0.694$&$0.504$&$0.775$\\\\

10&$\mathcal{D}_{PROP}$ & $\mathcal{D}_{GP}$&$0.774$&$0.798$&$0.836$&$0.595$&$0.829$\\
11&$\mathcal{D}_{PROP}$ & $\mathcal{D}_{3PM}$&$0.573$&$0.586$&$0.625$&$0.373$&$0.706$\\
12&$\mathcal{D}_{PROP}$ & $\mathcal{D}_{PROP}$&$0.712$&$0.763$&$0.761$&$0.517$&$0.796$\\\\

13&$\mathcal{D}_{GP3PM}$ & $\mathcal{D}_{GP3PM}$&$0.899$&$0.904$&$0.814$&$0.844$&$0.936$\\
14&$\mathcal{D}_{GP3PM}$ & $\mathcal{D}_{3PMGP}$&$0.082$&$0.067$&$0.080$&$0.061$&$0.076$\\
15&$\mathcal{D}_{3PMGP}$ & $\mathcal{D}_{GP3PM}$&$0.142$&$0.116$&$0.140$&$0.074$&$0.107$\\
16&$\mathcal{D}_{3PMGP}$ & $\mathcal{D}_{3PMGP}$&$0.829$&$0.856$&$0.851$&$0.626$&$0.910$\\

        \bottomrule
\end{tabular}}
\vspace{-1\baselineskip}
\end{table}

\paragraph{Similarity of \acp{apk}}
We begin by analyzing the impact on performance when training and testing on either the same or different markets. As shown in~\autoref{tab:data_market_results}, training and testing on the same market (rows 1--2) consistently yields higher \fone{}s compared to cross-market scenarios (rows 3--4). These results indicate notable differences in the characteristics of goodware and malware across markets.

This point is further underscored when examining the \fone of classifiers trained on samples from one market, but tested on an even distribution from mixed markets ($\mathcal{D}_{EVEN}$, rows 5--6). For example, training on $\mathcal{D}_{GP}$ and testing on $\mathcal{D}_{EVEN}$ yields an \fone significantly lower than those observed when testing on $\mathcal{D}_{GP}$ alone (row 1). Similar trends are observed for $\mathcal{D}_{3PM}$ and are consistent for all classifiers.

\paragraph{Combining markets for training/testing}
Prior work~\cite{pendlebury2019tesseract,barbero2022transcending,apigraph_paper} has adopted multi-market sampling. Here, we analyze training on mixed markets and testing on mixed or single markets. We find that, regardless of the proportion ($\mathcal{D}_{EVEN}$ or $\mathcal{D}_{PROP}$) of \googleplay to 3PM apps, the \fone when testing on multiple markets lies within the range of the \fone{}s when testing on the individual markets (rows 7--9), with $\mathcal{D}_{PROP}$ skewed towards the majority market (\googleplay in rows 10--12).

\paragraph{Goodware and malware from different markets}
Including goodware and malware from distinct markets ($\mathcal{D}_{GP3PM}$) leads to an unrealistically high \fone (rows 13 and 16). This aligns with findings from~\cite{dos_and_donts}, where classifiers inadvertently learn spurious correlations, distinguishing app origins rather than actual differences between goodware and malware. These results clearly demonstrate the influence of app markets on classifier performance and the risk of introducing biases.

\subsection{Recommendations} 
Arp et al.~\cite{dos_and_donts} showed that sampling malware and goodware from different markets can lead to spurious correlations. We extend this finding by showing that apps from different markets are inherently different (even if the class ratio remains consistent) and that training on one market does not transfer well to another. To ensure that goodware and malware distributions are not biased by their market source, one should sample both from the same market. To include multiple markets, one must sample from each separately and test on single markets to obtain best-case and worst-case estimates. 

\recommendationboxnew{Recommendation: \factorMarketsName}{The \factorMarketsName distribution \textbf{must} be kept consistent between goodware and malware to avoid spatial bias. When performing multi-market evaluation, single-market test performance \textbf{must} be reported to obtain the best- and worst-case outcomes.}

\section{Factor 4: VirusTotal Threshold}
\label{sec:factor-vtt}
\subsection{Hypothesis}

Although the \acl{vtt} is typically used to label a given dataset, it also acts as a sampling parameter: samples with a number of detections between one and the chosen \ac{vtt} value are excluded from the base population (these samples are usually called \textit{grayware}). This is particularly evident in \androzoo, where setting a \ac{vtt} of 4 excludes over 56\% of the non-benign population. For this reason, we posit that selecting different \ac{vtt} values will significantly affect the composition of malware in a dataset, consequently impacting the reported performance of a classifier. We focus on three representative \acp{vtt}, \ac{vtt}=15 based on \apigraph, \ac{vtt}=4 based on \transcend, and \ac{vtt}=2, based on the fact that previous literature demonstrated it is the lowest stable \ac{vtt} one can choose~\cite{zhu2020measuring}. 

\hypothesisbox{Hypothesis: \factorVTTName}{The choice of \ac{vtt} affects the distribution from which Android malware is sampled, as it acts as a sampling parameter that filters the base population, thereby affecting the reported performance of a classifier.}

\subsection{Impact}
In this section, we investigate the impact of \ac{vtt} on detection performance for both \apigraph and \transcend. To this end, we imitate the sampling performed by \apigraph and \transcend, varying only the \ac{vtt} values. More specifically, we sample new malware using \acp{vtt} of 2, 4 and 15 based on the amount of malware per month in \apigraph and \transcend. We imitate \transcend by filtering out samples that were not present on \androzoo after 2019. For \apigraph, we discard samples crawled by \androzoo after 2020 (the year \textit{APIGraph}~\cite{apigraph_paper} was published) and only include \acp{apk} found in VirusShare, \ac{vt}, and the AMD~\cite{amd_dataset} datasets. We perform each sampling three times for statistical significance. Further details regarding the imitation of the sampling of \apigraph and \transcend can be found in Appendix~\autoref{sec:appendix:datasets}.

\autoref{fig:vtt_hcc_fone}~shows the \fone of the \HCC classifier on the datasets sampled using different \ac{vtt} values. We include the performance plots for the remaining classifiers in Appendix~\autoref{sec:appendix:vtt}. We observe that resampling \apigraph with a \ac{vtt} of 4 does not statistically affect the AUT of an \HCC classifier when evaluating it on the complete evaluation window (\ie AUT=0.84 for the original \apigraph against an average AUT of 0.83$\pm$0.01). In contrast, using a \ac{vtt} of~2 does impact the reported performance, with the average AUT dropping to 0.77$\pm$0.03. However, we note that drift trends are accentuated in the last 18 months of our resampled datasets (particularly for \ac{vtt}=2), with the AUT dropping from 0.84 for the original \apigraph to~0.79$\pm$0.03 for \ac{vtt}=4 and to~0.68$\pm$0.05 for \ac{vtt}=2.

In the case of \transcend, employing a \ac{vtt} of 2 or 15 does not statistically affect the average yearly AUT (from 0.46 in the original dataset to 0.45$\pm$0.02 and 0.50$\pm$0.03). However, we notice that \transcend has only 75 samples in December, 2016, which affects evaluation metrics. When excluding this date, we find that datasets with \ac{vtt}=15 have an average AUT of 0.56$\pm$0.13 against an AUT of 0.49 for the original \transcend.

\begin{figure}[t]
    \begin{minipage}{0.49\linewidth}
        \centering
        \includegraphics[width=\textwidth]{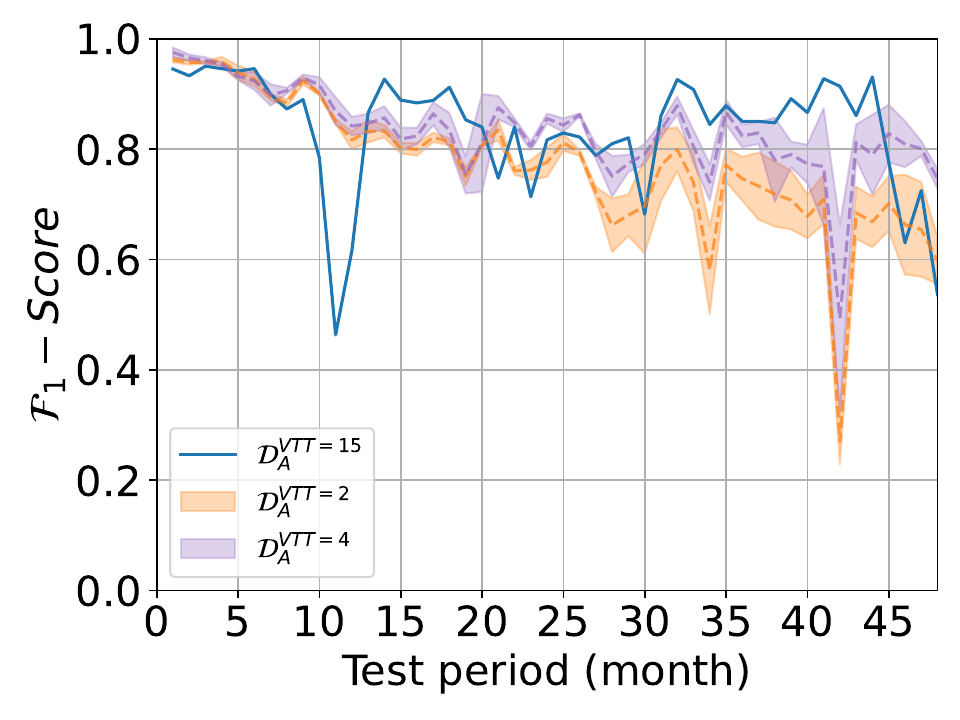}
        \subcaption{\apigraph}    
    \end{minipage}
    \hfill
    \begin{minipage}{0.49\linewidth}
        \centering
        \includegraphics[width=\linewidth]{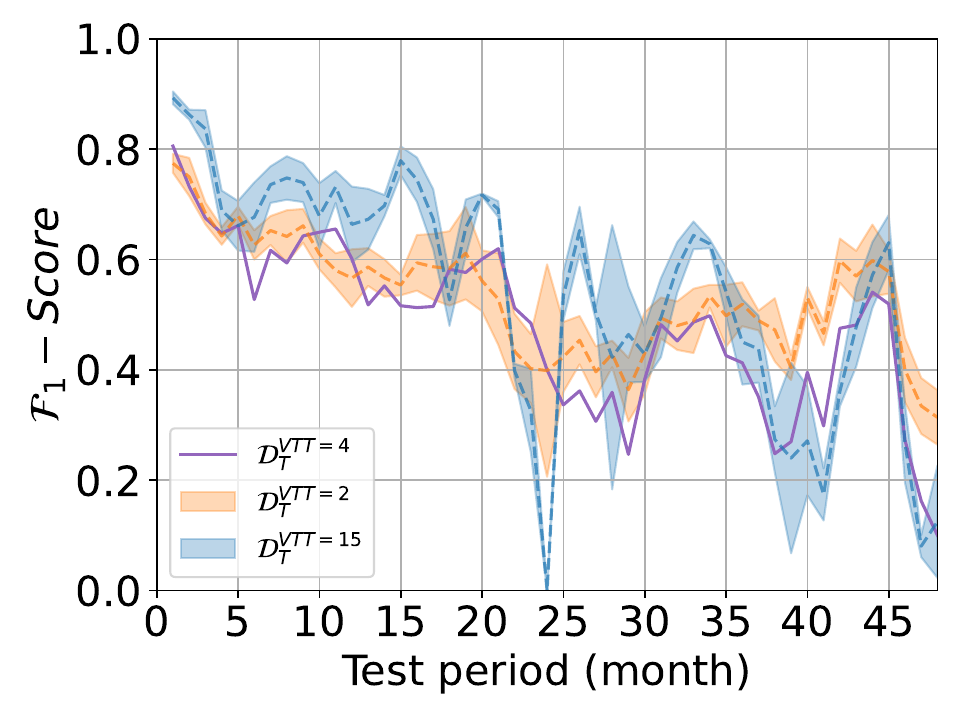}
        \subcaption{\transcend}
    \end{minipage}
    \caption{\textbf{\fonebold of \HCC classifier for \transcendbold and \apigraphbold sampled with a \ac{vtt}=2, \ac{vtt}=4, and \ac{vtt}=15}. All other classifiers are available in Appendix~\autoref{sec:appendix:vtt}.}
    \label{fig:vtt_hcc_fone}
\end{figure}

\subsection{Recommendations}
The choice of \ac{vtt} impacts the sampling process and may affect classification performance. 
Considering that a higher \ac{vtt} will filter out more samples, a lower \ac{vtt} will result in a more inclusive and therefore more representative dataset. 
Provided that a minimum of 2 detections is required to reduce noisy labels~\cite{zhu2020measuring}, we recommend using \ac{vtt}=2. 

\recommendationboxnew{\textbf{Recommendation: \factorVTTName}}{\ac{vtt}=2 \textbf{must} be used to sample from a distribution closer to the original one.}

\section{Factor 5: \factorSamplingName}\label{sec:factor-sampling-size}
\subsection{Hypothesis}
Evaluating classifier performance over the entire Android \ac{apk} population (or a surrogate such as \androzoo) is infeasible: \androzoo alone comprises more than 25.8 million samples spanning 12+ years. Conversely, arbitrarily small samples may not reflect the underlying population.

\apigraph and \transcend contain 241{,}611 and 259{,}230 samples, respectively, between 2014--2018, yet neither work justifies its chosen sample size. For instance, \apigraph samples 500 malware and nine times as much goodware per month, while \transcend uses varying monthly sizes in its early years before fixing the count at 500 malware and 5{,}000 goodware in 2017--2018. Although their statistical representativeness remains unclear, both datasets follow \tesseract domain constraints and best practices from prior work: $C1$ (temporal training consistency), $C2$ (consistent time windows for goodware and malware), and $C3$ (realistic 10\% malware-to-goodware testing ratios)~\cite{pendlebury2019tesseract}.

In contrast, Miranda et al.~\cite{miranda2022debiasing} propose a statistical framework that determines minimum sample sizes for representativeness using the classical \emph{margin of error} with finite population correction and a Bonferroni adjustment over multiple and security-unrelated characteristics (\eg binned APK size and release year)~\cite{miranda2022debiasing, casella2024statistical}. The key motivation for this design was that security-related characteristics (\eg cryptographic API calls) were considered hard to compute and, if incorporated, might introduce bias into the dataset. We refer to this framework as \dada.

However, certain domain-specific constraints should still be taken into account when determining the sample size. We posit that combining the uniform statistical sampling of \dada with domain-specific constraints from \tesseract leads to more consistent datasets in the same time frame.

\hypothesisbox{Hypothesis: \textbf{\factorSamplingName}}{Relying solely on non-security related characteristics is insufficient for constructing datasets appropriate for Android malware classification; security-specific constraints proposed in \tesseract~\cite{pendlebury2019tesseract} are necessary to ensure representative datasets.}

\subsection{Impact}
\paragraph{Adapting \dada}
\dada~\cite{miranda2022debiasing} originally estimates the minimal sample size using uniform sampling, \ie without distinguishing between malware and goodware, and only accounting for general application characteristics shared across both classes. We introduce two incremental modifications to \dada. First, we move from uniform to stratified sampling by applying \dada independently to malware and goodware (\textsc{Stratified}), ensuring statistically sufficient representation of both classes. Second, to align with constraint $C3$ from \tesseract, which suggests enforcing a 1:9 malware to goodware test ratio~\cite{pendlebury2019tesseract}, we compute the minimum sample size per class and then increase goodware to nine times the malware amount, yielding realistic malware-to-goodware ratios while preserving the same malware volume~\cite{pendlebury2019tesseract}. Moreover, notice that the malware-to-goodware ratio is a parameter of the algorithm and can therefore be changed to other values. This flexibility is required to address potential future changes in the percentage of malware in the wild.
We refer to the resulting procedure as \emph{Statistically representative, \tesseract-guided Application Sampling strategy} (\stas). This method combines the benefits of stratified sampling with realistic class ratios.

\paragraph{Sampling new datasets}
To assess the impact of uniform and stratified sampling, we construct nine datasets using \dada, stratified sampling, and \stas. 
Guided by our findings and recommendations in \autoref{sec:factor-timestamps}, \autoref{sec:factor-app-markets}, and \autoref{sec:factor-vtt}, we apply a \ac{vtt}=2 to capture more of the population, eliminate market-specific biases by restricting sampling to \googleplay, and use \googleplay \uploaddates as the most reliable timestamps. Following prior work showing that malware labels stabilize within one year~\cite{miller2016reviewer,zhu2020measuring}, we sample three years of data spanning 2021--2023. We apply the generic Android characteristics defined in \dada (\ie 4 values of APKs, 10 boolean permissions) and only change \ac{apk} release year to 2021--2023 (3 values).\footnote{To aid future works, we release \hypercube, the first version of the dataset sampled using \stas following the aforementioned parameters. Although \hypercube is in a fixed time frame, the \stas methodology can be easily used to sample newer datasets in the future. }

\begin{figure}
    \centering
    \includegraphics[width=0.60\linewidth]{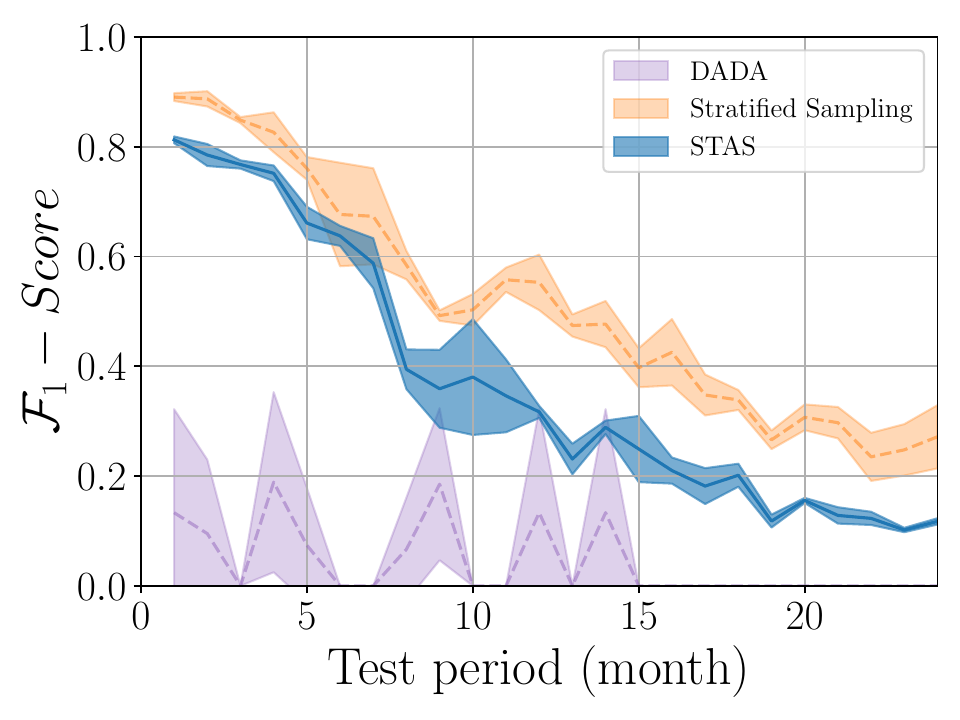}
    \caption{\textbf{\fonebold score for \HCC on datasets sampled using \dada, \textsc{Stratified} and \stas}. \dada alone is insufficient for creating Android malware datasets. \textsc{Stratified} inflates detection performance since it produces unrealistic class ratios (almost 50\% malware). \stas creates stable and realistic datasets. All other classifiers are available in Appendix~\autoref{sec:appendix:dataset-size}.}
    \label{fig:dada_vs_hypercube3_hcc}
\end{figure}

\begin{table}[t]
    \centering
    \scriptsize
    \caption{\textbf{Average A-AUT across three datasets sampled using \dada, \textsc{Stratified}, and \stas for 2021--2023}. Depending on the sampling strategy employed, the resulting dataset size changes, affecting the overall performance.}
    \begin{tabular}{l|rrr}
         & \multicolumn{3}{c}{\textbf{Average A-AUT across three datasets}} \\
         
         \textbf{Classifier} & \dada & \textsc{Stratified} &  \stas \\ \midrule
         \drebin & 0.09$\pm$0.03 & 0.59$\pm$0.01 & 0.43$\pm$0.00\\
         \deepdrebin & 0.01$\pm$0.01 & 0.61$\pm$0.01 & 0.44$\pm$0.01\\ 
         \malscan & 0.11$\pm$0.03 & 0.58$\pm$0.01 & 0.35$\pm$0.00 \\ 
         \ramda & 0.01$\pm$0.01 & 0.55$\pm$0.02 & 0.27$\pm$0.03\\ 
         \HCC & 0.03$\pm$0.03 & 0.60$\pm$0.02 & 0.46$\pm$0.00 \\ 
         \midrule
         Malware Size & 206 (0.78\%) & 19,645 (42\%) & 19,645 (10\%)\\
         Goodware Size & 26,741 (99.22\%) & 26,946 (58\%) & 176,805 (90\%)\\
         \midrule
         \textbf{Dataset size} & 26, 497 & 46,591 & 196,450\\ 
    \end{tabular}
    \label{tab:dada_stas_all}
\end{table}

\autoref{fig:dada_vs_hypercube3_hcc} shows the temporal performance of the three sampling approaches. A key distinction between \dada, a uniform sampling approach, versus the two stratified sampling approaches is that the latter guarantees a statistically sufficient number of both goodware and malware samples (\autoref{tab:dada_stas_all}). \dada yields a small dataset of 26,947 samples compared to 46,591 samples for stratified sampling without enforcing \tesseract's $C3$ constraint. However, the implications is that stratified sampling produces a relatively stable \fone, suggesting suitability for Android malware evaluation.

Both stratified sampling approaches show low variance across multiple executions, but \stas shows the overall lowest \fone. Although a lower \fone does not necessarily indicate better representativeness, it correlates with the findings of Liu et al.~\cite{liu2022explainable} and \tesseract~\cite{pendlebury2019tesseract}, where an imbalance of malware-to-goodware can inflate performance results. Moreover, following \tesseract's recommendation of enforcing $C3$ would result in a significantly larger dataset size of 196,450 samples. While this entails higher computational costs, \stas nevertheless represents the minimum sample size to construct a dataset that is both statistically representative and realistic. 

\subsection{Recommendations}
We show that combining temporal and spatial constraints with statistical sampling algorithms is necessary to estimate an adequate minimum number of samples required for a dataset to be representative of the real-world population.

\recommendationboxnew{Recommendation: \factorSamplingName}{Dataset sizes \textbf{must} be statistically representative of the population and follow domain-specific guidelines. \stas \textbf{may} be used to determine appropriate sizes for malware datasets.}

\begin{figure*}[t]
    \centering
    \includegraphics[width=1\linewidth]{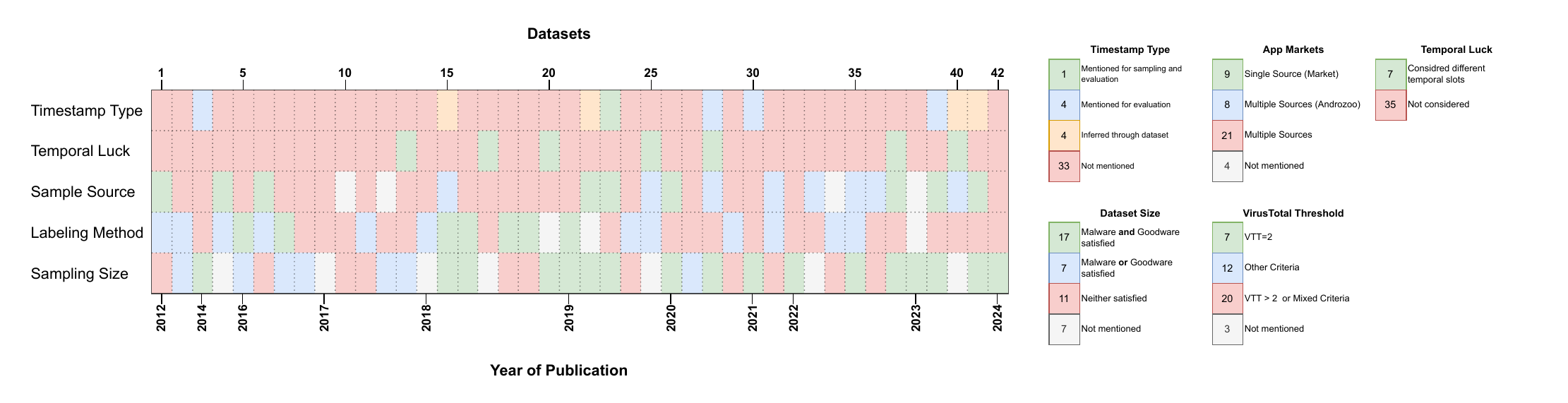}
    \caption{\textbf{Summary of the prevalence analysis of bias factors} in datasets used in the top-four security conferences and in the top-100 Google Scholar results. Of the 42 datasets we surveyed, only two datasets satisfy three of our recommendations~\cite{pendlebury2019tesseract, barbero2022transcending}, while all other datasets violate three or more. This suggests our factors are largely overlooked in previous work.}
    \label{fig:prevalence}
\end{figure*}

\section{Factor Prevalence Analysis}
\label{sec:prevalence}
To confirm whether the identified factors have been overlooked by the research community, we conduct a thorough analysis of existing Android malware datasets. 

We first compile a representative list of Android malware datasets used in previous work using a forward-citation methodology, \ie identifying papers that cite a given dataset. We begin with two of the earliest and most widely used datasets: Malgenome~\cite{zhou2012dissecting} and Drebin~\cite{arp2014drebin}, each with over 3,000 citations and published in a top security venue (IEEE S\&P 2012 and NDSS 2014, respectively). 

To narrow our scope, we initially include works published in one of the top-four security venues: USENIX Security, ACM CCS, IEEE S\&P, and NDSS. This helps us filter out papers that do not address security topics and focus on those that apply machine learning to security-related research problems. We also include relevant works outside these venues by querying Google Scholar for ``Android Malware Detection'' and selecting the top 100 results (as of July 10th, 2025).

This results in an initial list of 233 articles, which we individually review to identify datasets used for experimental evaluation. To ensure comprehensiveness of our survey, we perform forward-citations for every dataset that we have identified and filter for the top-four security venues. This not only informs us of the prevalence of each dataset, but also reduces the chance of missing relevant ones. 

In total, we analyze 527 Android malware detection papers and identify 42 datasets, which are explicitly used in 154 papers, including 35 from top-four security venues. Appendix~\autoref{sec:appendix:prevalence} details our methodology, dataset characteristics, and factor assessment criteria.

\autoref{fig:prevalence} summarizes the prevalence of the five factors across the 42 datasets. We observe that crucial curation practices are often neglected: 33 datasets ignored timestamp types for sampling (\autoref{sec:factor-timestamps}), which was shown to affect the distribution the sampled dataset represents; only 7 evaluated their dataset on different temporal splits; and over half mixed \ac{apk} sources. Labeling practices were similarly inconsistent, with 7 datasets using \ac{vtt}=2, 12 relying on other criteria, and 20 employed \ac{vtt}$>$2 or mixed methods. Although 17 datasets were statistically significant in size, none explicitly justified their chosen dataset size. Ultimately, only 2 datasets~\cite{pendlebury2019tesseract, barbero2022transcending} satisfied at least three of our recommendations, while the remaining 40 violated three or more.

Overall, our findings suggest that best practices defined in prior work have not clearly outlined and addressed the issues we have highlighted.

\section{Discussion}
\label{sec:discussion}

\daniel{TODO: General question - Do we want to sell our recommendations as generally applicable to all malware domains or not?}
While we extend the set of best practices for evaluating Android malware classifiers under spatio-temporal drift (\cite{dos_and_donts,pendlebury2019tesseract,rossow2012prudent,sommer2010outside,flood2024bad}), several open problems still exist. Here, we contextualize our findings and discuss related challenges.

\paragraph{Other malware domains}
While our study focuses on Android malware (as it is the only domain for which large-scale timestamped repositories of both benign and malicious apps are available), many of the challenges may be shared with other malware domains, such as Windows and PDFs. For example, getting a reliable timestamp source can be difficult, especially when there are no centralized app markets such as in Android. Jiang et al.~\cite{jiang2024benchmfc} recommend using \vtdates as the best estimate for Windows PE; however, we have shown how Third-Party Timestamps may be misaligned with the real publication date of software, at least for \googleplay. \factorTemporalLuckName and our recommendation remains relevant in other malware domains as well. Lastly, our algorithm for estimating the dataset size can be applied to other malware domains if the general population distribution is available.

\paragraph{Evolution of \tesseract $C3$~\cite{pendlebury2019tesseract}}
We observed that malware is not consistently $10\%$ across all months in \androzoo. However, obtaining a ``true'' malware ratio would require help from industrial partners. \tesseract~\cite{pendlebury2019tesseract} estimated the percentage of malware samples at $10\%$ in 2019, which may no longer be valid in the current landscape. For lack of a better source, assuming $10\%$ as a constraint allows for consistency in the sampled datasets. Future work should investigate the temporal evolution of realistic malware ratios.

\paragraph{K-fold evaluation} \tesseract explained that K-fold cross-validation is an upper bound estimation for performance in the absence of drift~\cite{pendlebury2019tesseract} and proposed AUT for time-aware evaluations~\cite{pendlebury2019tesseract}. In~\autoref{sec:factor-temporal-luck}, we argue that temporal evaluation should be performed with rolling-window splits to mitigate non-stationary ``lucky splits,'' similar to how K-fold avoids stationary lucky splits. Although this approach is standard for financial time-series forecasting~\cite{de2018advances}, its implications remain largely unexplored in the security domain. 

\paragraph{Family trends in a dataset} This work highlights the prevalence of re-emerged malware (\autoref{sec:factor-timestamps}) in the wild, which may impact the reported performance of a classifier. This is indirectly reflected in our discussion on ``lucky temporal splits,'' to which the A-AUT offers a mitigation (\autoref{sec:factor-temporal-luck}). 
Showing the family overlap in a dataset may help better understand the impact this may have on the underlying classifier. We acknowledge this requires a well-defined protocol to encourage sound comparisons; we highlight this as a current limitation of our approach, leaving this as future work.

\paragraph{Grayware}
Our work provides evidence that lower \vtthreshold values better represent the malware population. While we generally consider the problem of benign vs. non-benign, accounting for \textit{grayware} (\acp{apk} detected by at least one \acl{AV} but less than \vtthreshold), which are predominant in \androzoo, remains an open problem. Previous work~\cite{andow2016study} has attempted to categorize different grayware types in \googleplay. However, systematizing grayware also remains an open problem. 

\paragraph{Adversarial settings}
Adversarial attacks in the Android malware domain typically involve modifications of the malicious code that induce misclassifications during classification~\cite{deepdrebin,li2019adversarial}. While such settings are important for evaluating the robustness of classifiers, the focus of this work is on curating realistic datasets that capture natural distribution shift of Android malware, not adversarial benchmark datasets. Future research should further investigate how dataset biases may influence evaluations under adversarial conditions.
Similarly, real-world attackers have been using code obfuscation techniques and packing to evade malware detectors~\cite{elsersy2022rise}. Future work should investigate the impact of code packers on dataset construction.

\paragraph{Alternative sampling strategies}
We acknowledge that our statistical-based down-sampling may unintentionally under-represent rare malware families, as these are statistically less prevalent in the population. Consequently, datasets sampled using our approach may be dominated by widespread families while excluding those with fewer samples. Addressing this limitation is non-trivial: obtaining reliable family labels for entire malware populations is impractical due to VirusTotal API rate limits and inconsistent labeling across antivirus vendors. We leave the exploration of family-aware sampling strategies to future work.

\section{Related Work}
\label{sec:related_works}

In this section, we cover the research most closely related to our work. We present both seminal research papers tackling the wider problem of achieving fair evaluations in \ac{ML}-based computer security research and more directly comparable publications on dataset sampling and debiasing.

\paragraph{Guidelines for realistic \ac{ML} evaluations in computer security} Our research aligns with efforts to establish guidelines for realistic evaluations in security-related applications. Sommer and Paxson~\cite{sommer2010outside} highlighted key challenges in applying \ac{ML} to network intrusion detection systems (NIDS), identifying fundamental issues that persist in this domain. Rossow et al.~\cite{rossow2012prudent} and Van der Kouwe et al.~\cite{van2019sok} further analyzed common shortcomings and best practices in security system evaluations. However, these works primarily identify challenges and methodological pitfalls without quantifying their impact on classifier performance. Arp et al.~\cite{dos_and_donts} extended this line of work by demonstrating the effects of these pitfalls in realistic scenarios, although with a focus on general cybersecurity.
In contrast, our work investigates the specific properties and risk factors during dataset creation for the Android malware domain that lead to pitfalls identified in prior studies. 

\paragraph{Temporal evaluations} Other relevant work includes the domain of temporal evaluation, primarily treated by Allix et al.~\cite{allix2015your} and Miller et al.~\cite{miller2016reviewer}. Recently, Pendlebury et al.~\cite{pendlebury2019tesseract} proposed \tesseract as a framework to measure the impact of performance decay over time. We expand on \tesseract by proposing a new evaluation guideline in \factorTemporalLuckName. Additionally, we address the creation timestamps validity by experimentally showing the impact of inconsistencies and providing practical recommendations for researchers (\autoref{sec:factor-timestamps}). 

\paragraph{Realistic datasets for ML-based cybersecurity} Prior work also focused on the problem of realistic and benchmark datasets for ML-based cybersecurity. Sommer and Paxon~\cite{sommer2010outside} highlighted the challenges of creating realistic network intrusion detection datasets, mostly due to privacy concerns; attempts to create datasets artificially, such as DARPA98-99\cite{lippmann20001999} and KDD99\cite{kdd_cup}, failed because they introduced artifacts~\cite{mahoney2003analysis, engelen2021troubleshooting, mchugh2000testing}. In the mobile malware domain, Haque et al. proposed a new dataset, LAMBDA~\cite{haque2025lamda}, which contains over one million \acp{apk} collected across 12 years, using crawl dates and \vtthreshold{}=4. However, LAMBDA does not resolve the biases we identify: \stas is a sampling strategy designed to help researchers build statistically sound and bias-aware datasets for any past or future time window, rather than simply aggregating a large corpus of samples. 

Similarly, Jian et al.~\cite{jiang2024benchmfc} created a benchmark dataset for Windows malware by sampling from VirusShare, following a similar dataset curation pipeline to the one we presented. However, we focus on factors related to representativeness, such as sample size and markets, rather than the quality of family labels. Ceschin et al.~\cite{ceschin2024machine} also presented a preliminary discussion on dataset size and markets but lacks empirical evaluation or concrete recommendations, while we provide both. Notably, they call for better dataset practices---our work directly answers their call. Recently, Flood et al.~\cite{flood2024bad} identified six poor practices in NIDS-specific datasets, including poor data diversity, highly dependent features, unclear ground truths, traffic collapse, artificial diversity, and wrong labels. Instead of focusing on the limitations of network datasets, we focus our attention on the Android malware domain, where the presence of \androzoo~\cite{androzoourl,allix2016androzoo,alecci2024androzoo} allows us to build more realistic datasets. Furthermore, our factors are directly applicable during the sampling process, helping to prevent the identified ``bad'' choices in \cite{flood2024bad}. Ge et al.~\cite{ge2021impact} analyzed three dataset factors (\ie class imbalance, quality, and timelines) by examining three state-of-the-art datasets. We differ from this work as we not only show how our sampling factors can affect the sampled distribution, but also give actionable recommendations that researchers can follow to reduce the biases highlighted in \cite{ge2021impact}. Additionally, we do not limit our work to only two state-of-the-art datasets, but instead include the wider Android population in our experimental analysis. In conclusion, we identify five factors that can affect realistic evaluations and that can be controlled with actionable recommendations (\autoref{sec:preliminary-observations}).

\paragraph{Research on Android malware sampling} Prior work has focused on the creation of representative Android benchmark datasets. Miranda et al. proposed \dada~\cite{miranda2022debiasing}, a two-step sampling approach for debiasing Android malware datasets, which combines statistical methods with general \ac{apk} characteristics. However, in~\autoref{sec:factor-sampling-size}, we have shown that \dada lacks critical domain-specific knowledge. In particular, it violates $C3$ of \cite{pendlebury2019tesseract} and P1 of \cite{dos_and_donts}. Sun et al.~\cite{sun2024temporal} proposed a dataset restructuring algorithm that unrealistically requires family information and applies it to \cite{malnet}. However, it does not correct MalNet's severe spatial bias (only about 6\% goodware). A recent work~\cite{thirumuruganathan2024detecting} proposed an approach to detect and mitigate sampling bias between a labeled training and an unlabeled testing dataset, based on domain discrimination. In contrast, our work questions how well datasets represent the real-world population, rather than fitting a classifier on two divergent datasets.

\section{Conclusion}

Motivated by the stark performance difference of Android malware classifiers on the \textit{APIGraph} and \textit{Transcendent} datasets, we reflected on dataset curation parameters and identified five factors overlooked by previous research, which can lead to unrealistic datasets: \factorTimestampName, \factorTemporalLuckName, \factorMarketsName, \factorVTTName, and \factorSamplingName. 
We have shown their impact on dataset composition (via family overlap) and on detection performance of five state-of-the-art classifiers. Since these factors are deeply interconnected (\eg modifying one will affect the others), they need a cohesive approach: for each bias factor, we propose actionable recommendations to cancel their impact; we propose A-AUT as an evaluation metric to address the impact of lucky temporal splits; we then consolidate our findings in proposing \stas, a statistically guided \tesseract-constrained sampling strategy. 

To encourage future research to exercise greater caution when selecting parameters during dataset curation, we release code for the \stas strategy and the hashes of \hypercube, a dataset sampled for 2021--2023 following our guidelines.

Although curating a static benchmark dataset may not be practical for Android malware classification due to natural distribution shift, following principled dataset curation practices is essential. Doing so will enable fairer, more reliable, and ultimately more trustworthy evaluations in this domain.

\section*{Acknowledgment}
This research was partially supported by: the UK EPSRC Grant EP/X015971/2; 
Google ASPIRE and GARA Awards; and
the Vienna Science and Technology Fund (WWTF) through the BREADS project (10.47379/VRG23011).

\bibliographystyle{plain}
\bibliography{ref}

\appendix
\section{Appendix}
\subsection{Datasets used in Evaluations}
\label{sec:appendix:datasets}
\paragraph{\textit{APIGraph} and \textit{Transcendent} Datasets}
Our hypotheses originate from a motivational example concerning two state-of-the-art datasets, \textit{APIGraph}~\cite{apigraph_paper} and \textit{Transcendent}~\cite{barbero2022transcending}. Of the 527 papers we reviewed in~\autoref{sec:prevalence}, we found 2 papers published in top conferences that use them. We also noticed that many of the papers sampled their own dataset, suggesting a lack of consistency in the datasets used in the scientific literature. 

\paragraph{App Markets dataset}
In~\autoref{sec:factor-app-markets} we collected 15,000 goodware and 15,000 malware samples for each \googleplay and third-party app markets through \androzoo, resulting in a total of 60,000 samples. We then combined them to create 6 different dataset configurations detailed in \autoref{tab:data_market_split}.

\begin{table*}[h]
\scriptsize
    \caption{\textbf{Experimental configurations for the analysis of market sources.} The configurations include sampling solely from (GP) or the Third Party Markets (3PM), from both evenly, from both but with more from \googleplay, goodware from \googleplay and malware from 3PM, and vice versa.}
    \label{tab:data_market_split}
    \centering
    \begin{tabular}{l|rrrr|rrrr}
    \multirow{2}{*}{\textbf{Configuration}} & \multicolumn{4}{c}{\textit{\textbf{Train set}}} & \multicolumn{4}{c}{\textit{\textbf{Test set}}}\\
         & \multicolumn{2}{c}{Goodware} & \multicolumn{2}{c}{Malware} & \multicolumn{2}{c}{Goodware} & \multicolumn{2}{c}{Malware}\\
        & GooglePlay & 3PM & GooglePlay & 3PM & GooglePlay & 3PM & GooglePlay & 3PM\\
        \midrule
        $\mathcal{D}_{GP}$   & 10,000 & 0 & 10,000 & 0 & 4,500 & 0 & 500 & 0\\
        $\mathcal{D}_{3PM}$      & 0 & 10,000 & 0 & 10,000 & 0 & 4,500 & 0 & 500\\
        $\mathcal{D}_{EVEN}$         & 5,000 & 5,000 & 5,000 & 5,000 & 2,250 & 2,250 & 250 & 250\\
        $\mathcal{D}_{PROP}$ & 8,000 & 2,000 & 8,000 & 2,000 & 3,600 & 900 & 4,000 & 100\\ 
        
        $\mathcal{D}_{GP3PM}$ & 10,000 & 0 & 0 & 10,000 & 4,500 & 0 & 0 & 500\\
        $\mathcal{D}_{3PMGP}$ & 0 & 10,000 & 10,000 & 0 & 0 & 4,500 & 500 & 0\\
        \bottomrule
    \end{tabular}
\end{table*}

\paragraph{\hypercube}
The \hypercube dataset was sampled using \stas, a statistically guided \tesseract constrained sampling strategy described in~\autoref{sec:factor-sampling-size}. We used \stas to determine the amount of malware and goodware per month required for a representative dataset. We chose a recent time frame of 2021--2023, as it was shown in prior work that labels tend to stabilize after one year~\cite{miller2016reviewer,zhu2020measuring}. The dataset employs \ac{vtt}=2 (c.f. \autoref{sec:factor-vtt}), includes samples from \googleplay only (c.f. \autoref{sec:factor-app-markets}), and uses \googleplay \uploaddates for both sampling and evaluation (c.f. \autoref{sec:factor-timestamps}). In \autoref{sec:factor-sampling-size}, we sampled using \stas three times. \hypercube refers to the first version, which we release along with the code for \stas.

\paragraph{Imitating APIGraph}
In \autoref{sec:factor-timestamps} and \autoref{sec:factor-vtt}, we attempt to imitate the sampling in \apigraph by sampling \acp{apk} from \androzoo that was collected from VirusShare, VirusTotal, and the AMD dataset. We filtered out samples with a crawl date of 2020 (same publication year as \apigraph), as these would not have been in \androzoo when \apigraph was sampled. We kept the amount of malware and goodware per month exactly the same, so that the dataset size and the class ratio would be exactly the same as the original \apigraph. We used \dexdates for sampling as we were unable to obtain \vtdates for all samples on \androzoo. We used \ac{vtt}{}=15 (except in~\autoref{sec:factor-vtt}, where \ac{vtt} values were changed to 2 and 4, respectively).

\paragraph{Imitating Transcendent}
In \autoref{sec:timestamps-impact} and \autoref{sec:factor-vtt}, we attempt to imitate the sampling in \transcend by sampling \acp{apk} from \androzoo. Notice that \transcend was constructed in two different moments, with the first segment of the dataset (2014--2016) sampled in mid 2017 and the second part (2017--2018) collected in early 2019. We imitated this by filtering out samples using crawl dates after mid 2017 for the first part and early 2019 for the second part. We sampled from all markets in \androzoo for both malware and goodware, used \dexdates for sampling, and used \ac{vtt}{}=4 (except in ~\autoref{sec:factor-vtt}, \ac{vtt} values were changed to 2 and 15, respectively).

\subsection{Classifier Details}
\label{sec:appendix:classifier-details}
In addition to releasing the code of our approach, we also detail here the parameters used in all the classifiers mentioned in~\autoref{sec:experimental-setup} and the time required by each for training and inference. All training and evaluations were performed on an Intel(R) Core(TM) Ultra 9 185H CPU with 16 cores and a GeForce RTX 4070 GPU, using \transcend 2014 for training and \transcend 2015-2018 for inference. 
Note that we performed feature extraction for all three feature spaces before any experiment; therefore, the reported times do not include it.

For \drebin, we use a LinearSVM with the hyper-parameter $C=1$, which has been found to be representative for a variety of scenarios~\cite{pendlebury2019tesseract,arp2014drebin,barbero2022transcending,chen2023continuous,demontis2017yes}. A LinearSVM is a classical supervised ML algorithm that aims to find a hyperplane that separates data points belonging to the different classes (\ie goodware and malware). Both training and inference are relatively fast, with data preprocessing, feature reduction, and model fitting taking $\approx$122 seconds for one year of data, while inference took $\approx$346 seconds. 

\deepdrebin is implemented as a Multi Layer Perceptron (MLP) with two hidden and densely connected layers composed of 200 neurons each and an output layer with two neurons; hidden layers use the Rectified Linear Unit activation function, whereas the output neurons employ the softmax function. Training (including data preprocessing, feature reduction, and model fitting) took $\approx$267 seconds for 20 epochs, while inference took $\approx$263 seconds.

\malscan~\cite{malscan_paper} uses social-network-based centrality analysis to extract relevant features from function-call graphs in \acp{apk}. In this paper, we have used the configuration with degree centrality and a Random Forest Classifier, with the hyper-parameter $n_{estimators} = 100$. Training took $\approx$232 seconds but required up to 70GB of memory, whilst inference took $\approx$434 seconds. 

\ramda~\cite{ramda} employs a subset of the Drebin feature space (which includes permissions, intent actions, and sensitive API calls) to represent \acp{apk} and combines a variational autoencoder (VAE) with an MLP; in particular, the compressed representation learned by the VAE is fed into the MLP. A sample is classified as malware if either the reconstruction error of the VAE is over a certain threshold or the MLP outputs the corresponding label. In this paper, we used the same configuration as in the original work, setting $\lambda_1 = 10$, $\lambda_2 = 1$ and $\lambda_3 = 10$. Training took $\approx$1,034 seconds for 50 epochs of VAE training and 50 epochs of MLP training, while inference took $\approx$404 seconds.

\HCC~\cite{chen2023continuous} employs the Drebin feature space with a low-variance feature reduction ($0.1\%$). It combines an encoder, trained through contrastive hierarchical learning and an MLP. In this work, we train the model for up to 40 epochs, as we did not observe any performance gains past this point. Training (including data preprocessing, feature reduction, and model fitting) took $\approx$1,459 seconds, while inference took $\approx$433.

\subsection{Family Overlap Metric}
\label{sec:appendix:family-overlap-formula}

\paragraph{Family Overlap Metric}
In our analysis, we are interested in describing the \textit{temporal distribution shift} of the malware population within one dataset without having to rely on any classifier's performance. Therefore, we introduce \textit{family overlap} ($\Phi$), a metric that models the shift in malware distribution within a dataset. Shown previously in \autoref{fig:apigraph_transcendent_family_overlap}, it allows us to obtain more general insights on the dataset, without relying on the representation or the classifier employed. From this, we can observe how certain sampling parameters or algorithms influence the resulting dataset composition.

While methods quantifying the difference of distributions exist (\eg the Kullback-Leibler divergence), they typically require a representation space to operate in. In contrast, we designed the family overlap metric from domain knowledge to be representation-independent and interpretable.

The \textit{family overlap} ($\Phi$) measures the percentage of samples in a dataset $\mathcal{D}$ belonging to a family that was already known in a different dataset $\mathcal{D}_\text{ref}$.

\begin{equation}\label{eq:family_overlap}
    \Phi(\mathcal{D}, \mathcal{D}_{\text{ref}}) = 
    \frac{
        \left| \left\{ (x, y) \in \mathcal{D} \;\middle|\; \exists\, (x', y') \in \mathcal{D}_{\text{ref}} \text{ s.t. } y = y' \right\} \right|
    }{|\mathcal{D}|}.
\end{equation}
Here, $(x,y)$ are tuples of \acp{apk} and their malware family label. Unless specified otherwise, we use the training split of any given dataset as the reference dataset $\mathcal{D}_{\text{ref}}$ and some testing split for which to calculate the family overlap as $\mathcal{D}$.

\noindent
$\Phi$ is helpful for measuring and visualizing malware trends over time, as it describes the percentage of malware belonging to families that already existed in the past. 
Because it provides a compact representation of the distribution shift, it is particularly helpful for comparing the effects of different sampling strategies and is therefore widely used throughout the paper.

\subsection{Dynamic Time Warping}
Dynamic Time Warping (DTW) is an algorithm that measures the similarity between two time series. DTW tries to ``warp'' one time series to align it with the other; hence, it assumes that two time sequences are similar but ``out of phase.'' For this reason, we apply it in~\ref{sec:timestamps-impact} to compare the time distributions of samples when using different timestamp types, as these may just be misaligned; for example, the \vtdate time distribution may be in principle identical in shape to that of \googleplay \uploaddates, just shifted forward.

More specifically, DTW is a dynamic programming technique that attempts to minimize the total accumulated distance between points belonging to the two sequences. It achieves this by re-aligning the two sequences. Once the optimal solution is found, the total accumulated cost describes the dissimilarity between the two sequences.

\subsection{Temporal Luck}
\label{sec:appendix:tempora_luck_example}

\paragraph{A-AUT for smaller evaluation windows} When $n_E < n_T$, the evaluation window is smaller than the training window. To avoid gaps in temporal coverage, we advance by $n_E$ at each step, resulting in contiguous evaluation sets while training sets overlap:
\begin{align}
    T &:= \{D_{in_E:in_E+n_T-1}\}_{i=0}^{k-1} \\
    E &:= \{D_{n_T+in_E:n_T+(i+1)n_E-1}\}_{i=0}^{k-1} \\
    k &= \lfloor(|D| - n_T) / n_E\rfloor.
\end{align}

\paragraph{\factorTemporalLuckName Example}
We here provide an example of a dataset split according to \factorTemporalLuckName, as a useful overview to the A-AUT introduced in~\autoref{sec:factor-temporal-luck}. 

A dataset $\mathcal{D}$ containing data from January 2014 to December 2019 would be divided into the following dataset splits when using $n_T=n_E=12$ (months): 

\begin{multline}
    T := \{D_{2014}, D_{2015}, D_{2016}, D_{2017}, D_{2018}\} \\
    E := \{D_{2015}, D_{2016}, D_{2017}, D_{2018}, D_{2019}\}.
\label{eq:example_time_split_1}
\end{multline}

\begin{figure}[t!]
    \centering
    \begin{minipage}{\linewidth}
        \includegraphics[width=0.49\textwidth]{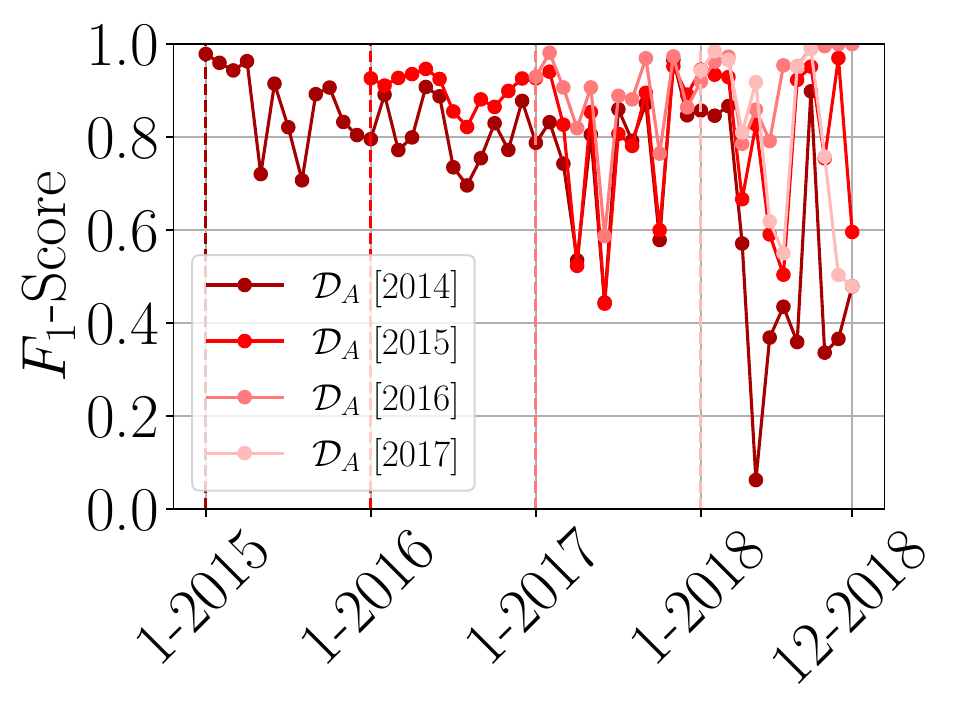}
        \includegraphics[width=0.49\textwidth]{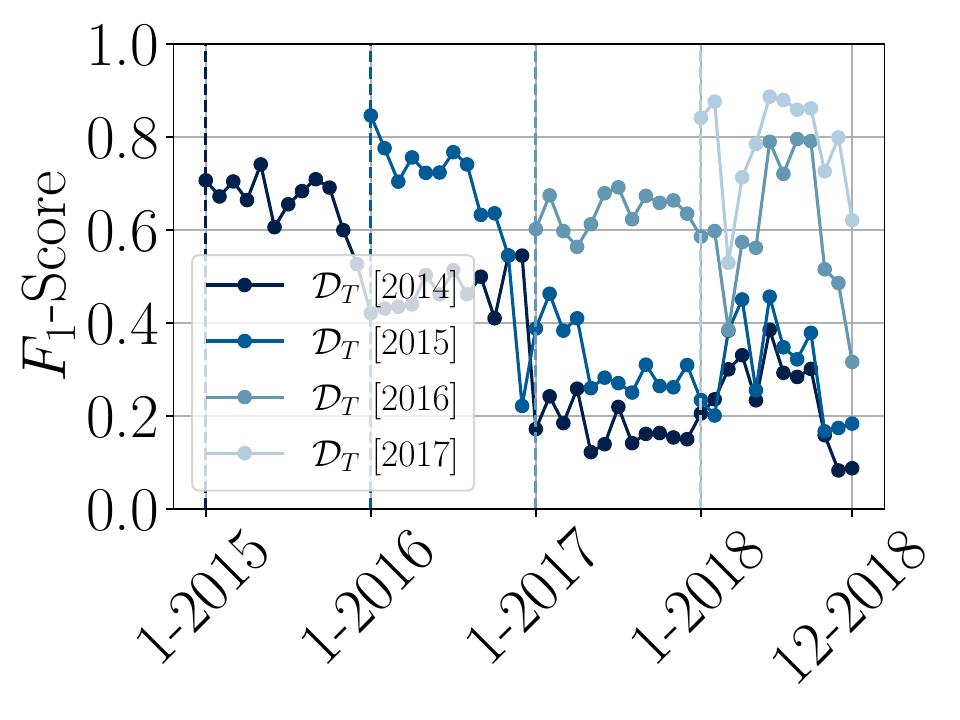}
        \subcaption{\drebin}
    \end{minipage}
    
    \begin{minipage}{\linewidth}
        \includegraphics[width=0.49\textwidth]{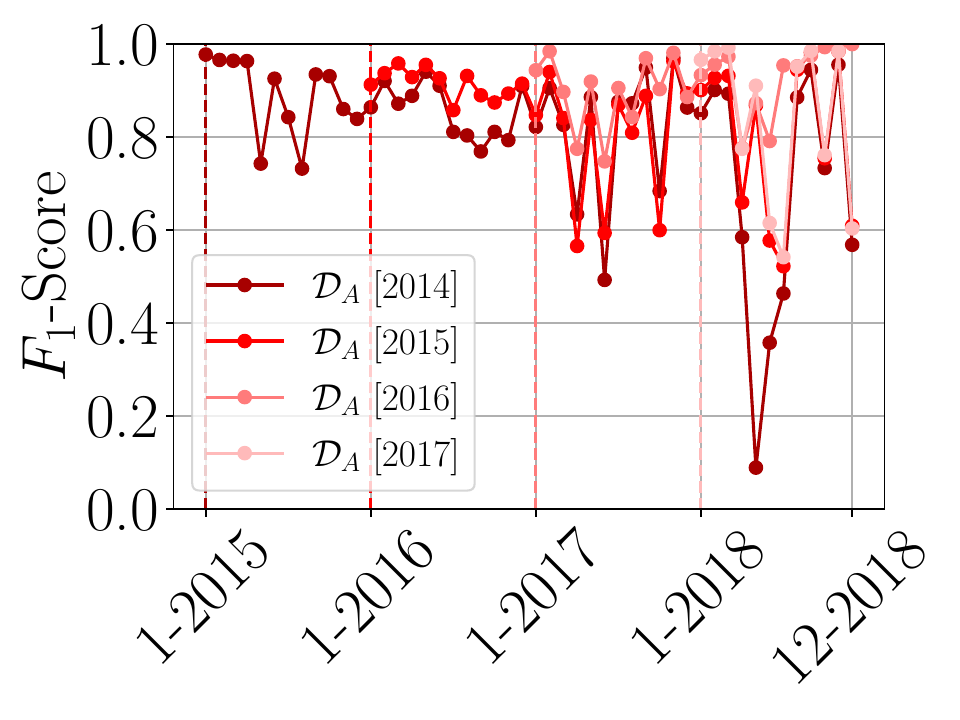}
        \includegraphics[width=0.49\textwidth]{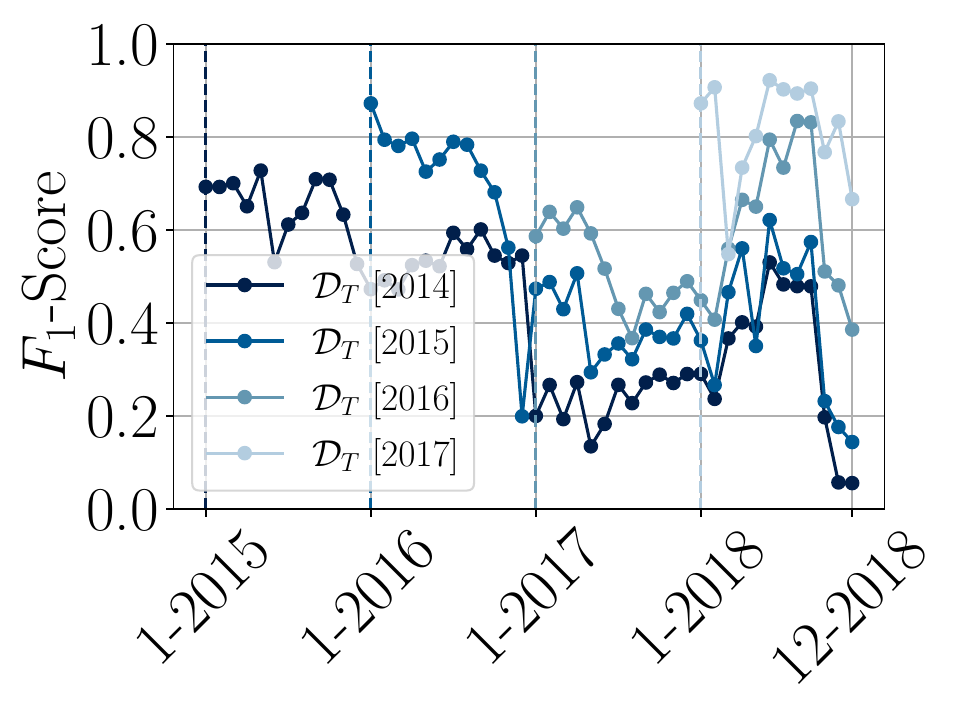}
        \subcaption{\deepdrebin}
    \end{minipage}

    \begin{minipage}{\linewidth}
        \includegraphics[width=0.49\textwidth]{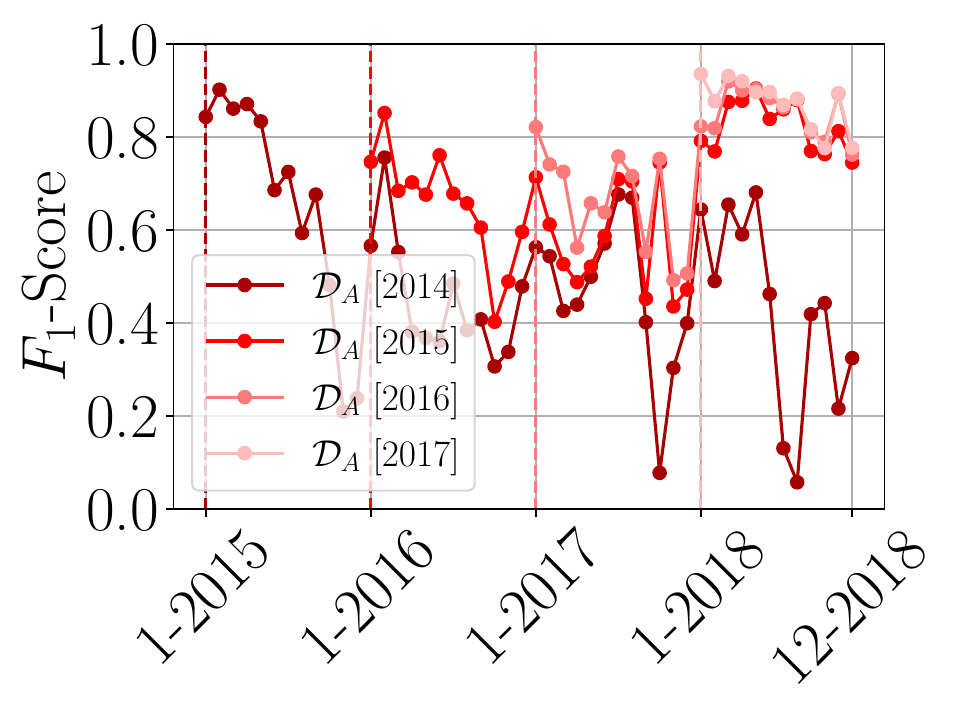}
        \includegraphics[width=0.49\textwidth]{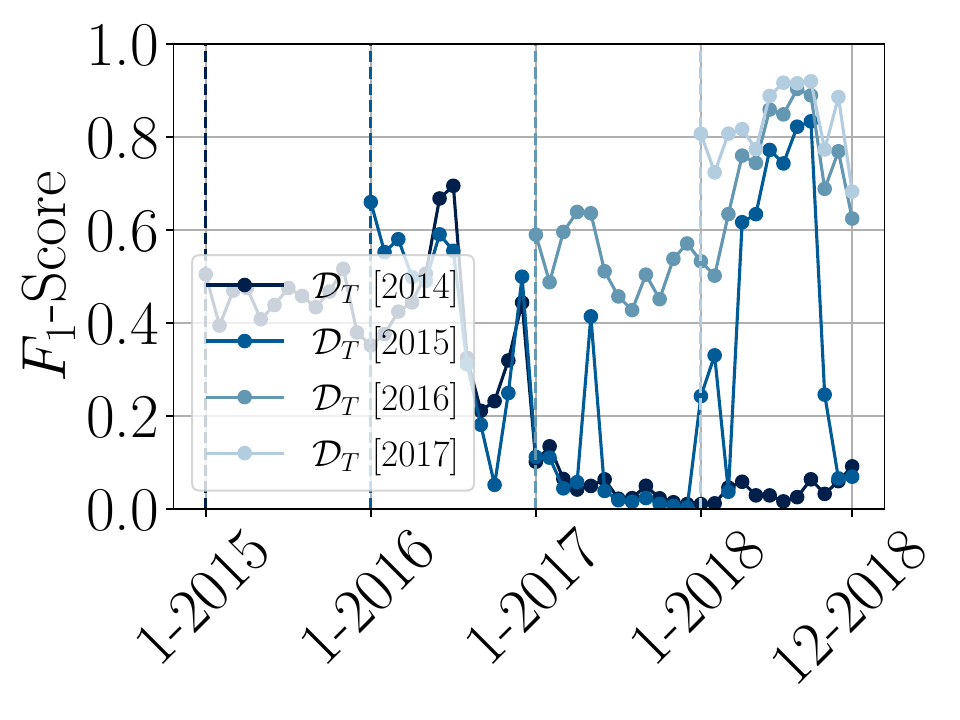}
        \subcaption{\malscan}
    \end{minipage}

    \begin{minipage}{\linewidth}
        \includegraphics[width=0.49\textwidth]{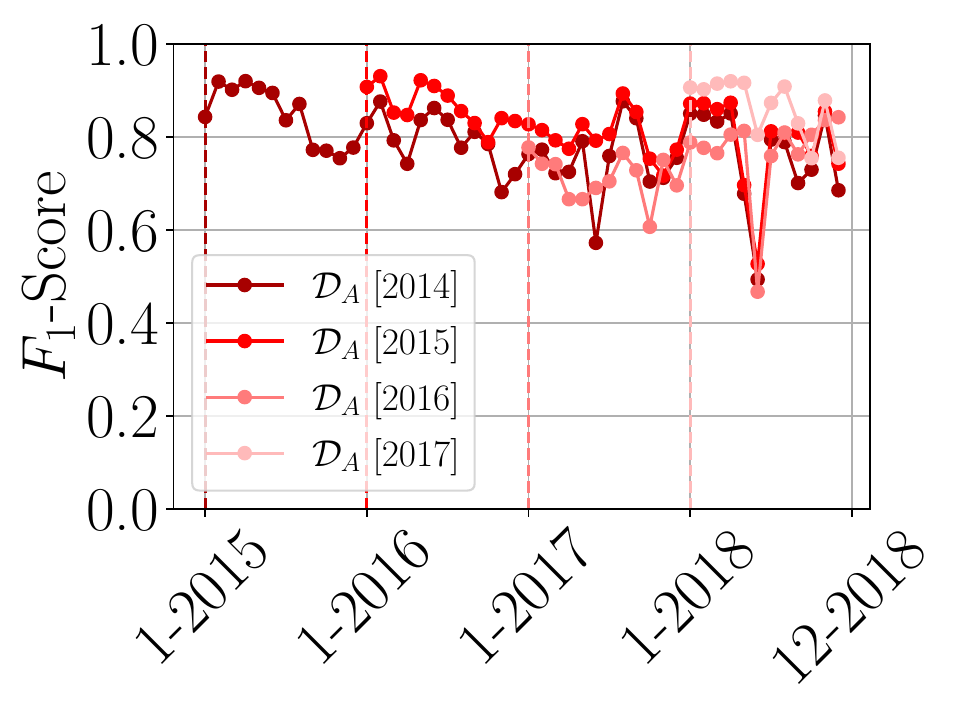}
        \includegraphics[width=0.49\textwidth]{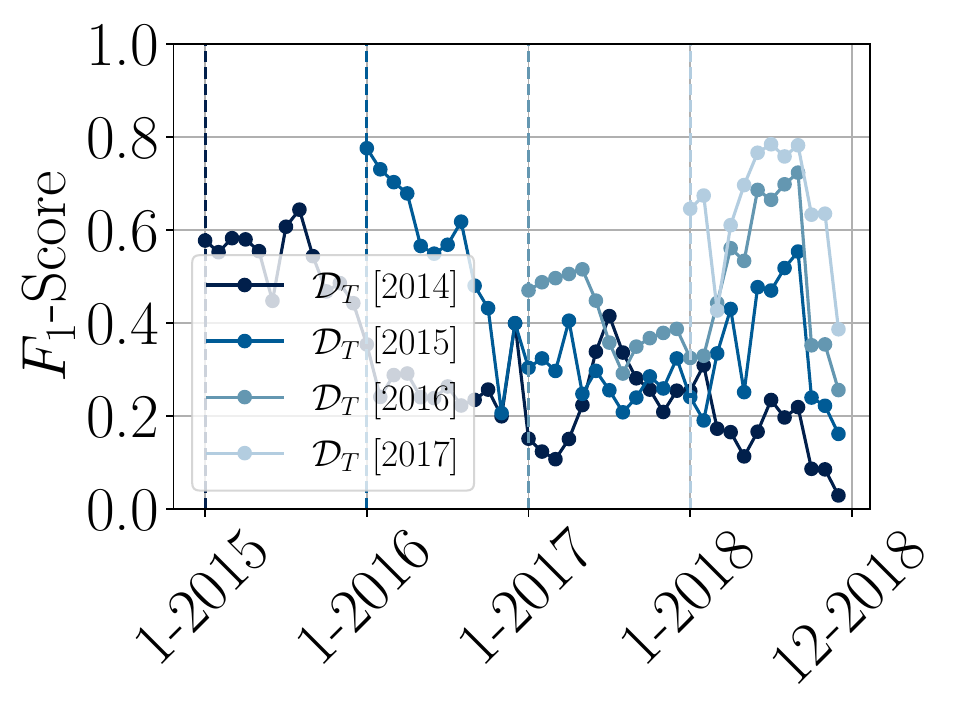}
        \subcaption{\ramda}
    \end{minipage}
    
    \begin{minipage}{\linewidth}
        \includegraphics[width=0.49\textwidth]{images/temporal_luck/motivational_HCC_APIGraph_plot.pdf}
        \includegraphics[width=0.49\textwidth]{images/temporal_luck/motivational_HCC_Transcendent_plot.pdf}
        \subcaption{\HCC}
    \end{minipage}
    
    \caption{\textbf{Impact of \factorTemporalLuckName on performance.} Each plot refers to a specific model and dataset. Each line indicates the \fone of a model trained on a different year (between 2014 and 2017). It can be seen that training on different years within the same dataset can yield different performance profiles. The issue is evident on \transcend, but also present in \apigraph.}
    \label{fig:temporal_luck_motivation}
\end{figure}

When $n_T\neq n_E$, for example $n_T=12$ and $n_E=24$ (months), $\mathcal{D}$ would be split into:
\begin{multline}
    T := \{D_{2014}, D_{2015}, D_{2016}, D_{2017}\} \\
    E := \{D_{2015:2016}, D_{2016:2017}, D_{2017:2018}, D_{2018:2019}\}.
\label{eq:example_time_split_2}
\end{multline}
\paragraph{Additional Temporal Luck results}
In~\autoref{sec:factor-temporal-luck}, we showed the effect of having different training and testing slots. We provide additional results for \drebin, \deepdrebin, \malscan and \ramda in~\autoref{fig:temporal_luck_motivation}. 

\subsection{VirusTotal Threshold}
\label{sec:appendix:vtt}
\begin{figure}[t]
    \begin{minipage}{\linewidth}
        \centering
        \includegraphics[width=0.49\textwidth]{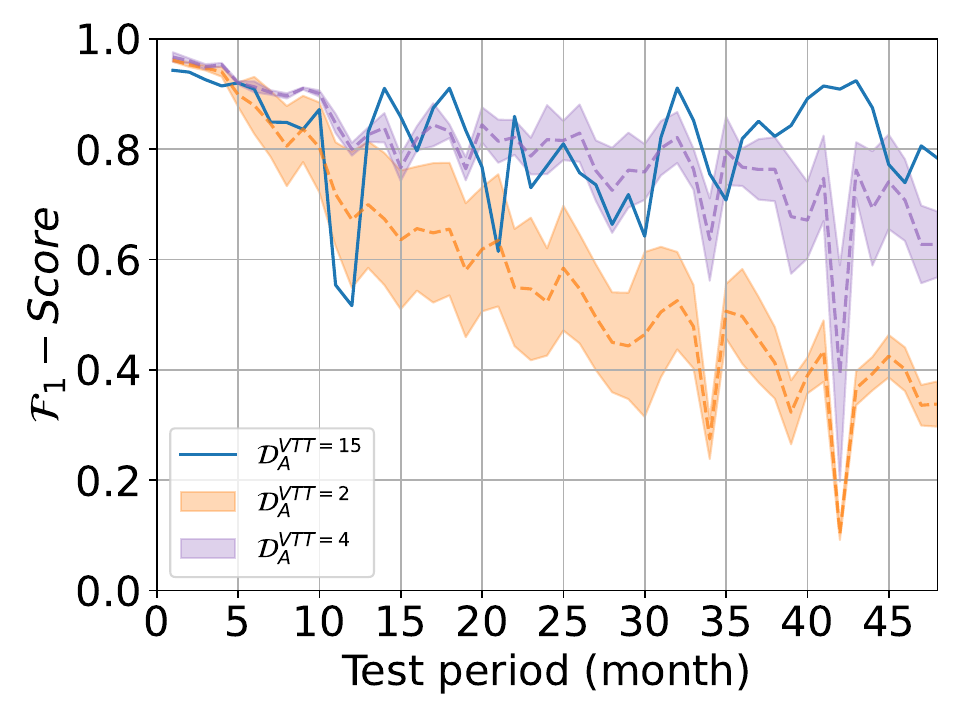}
        \includegraphics[width=0.49\textwidth]{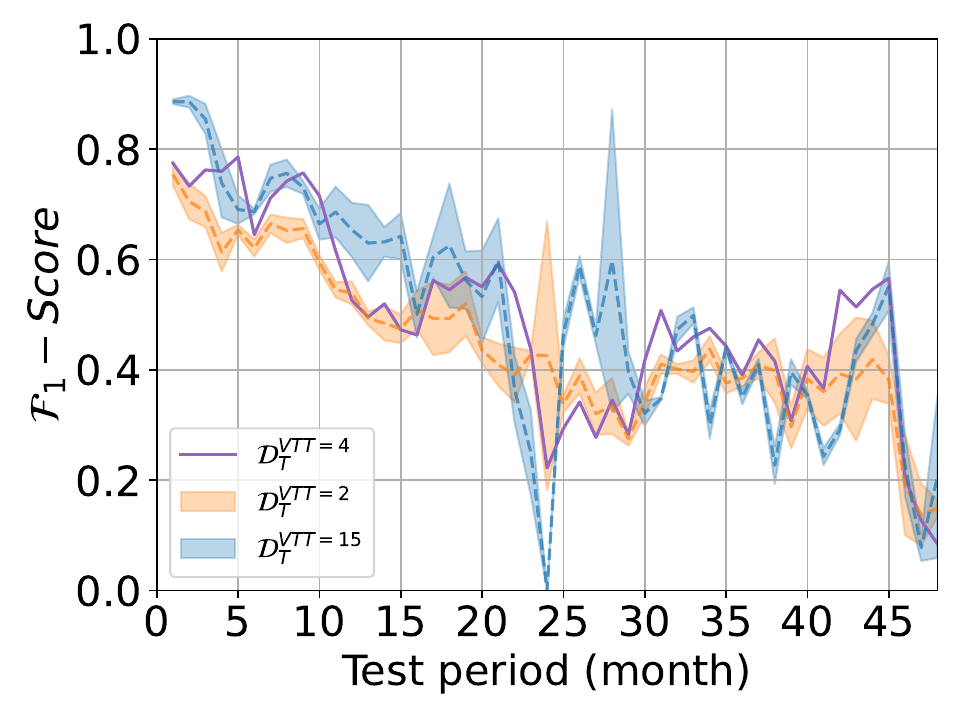}
        \subcaption{\drebin}    
    \end{minipage}
    \hfill
    \begin{minipage}{\linewidth}
        \centering
        \includegraphics[width=0.49\textwidth]{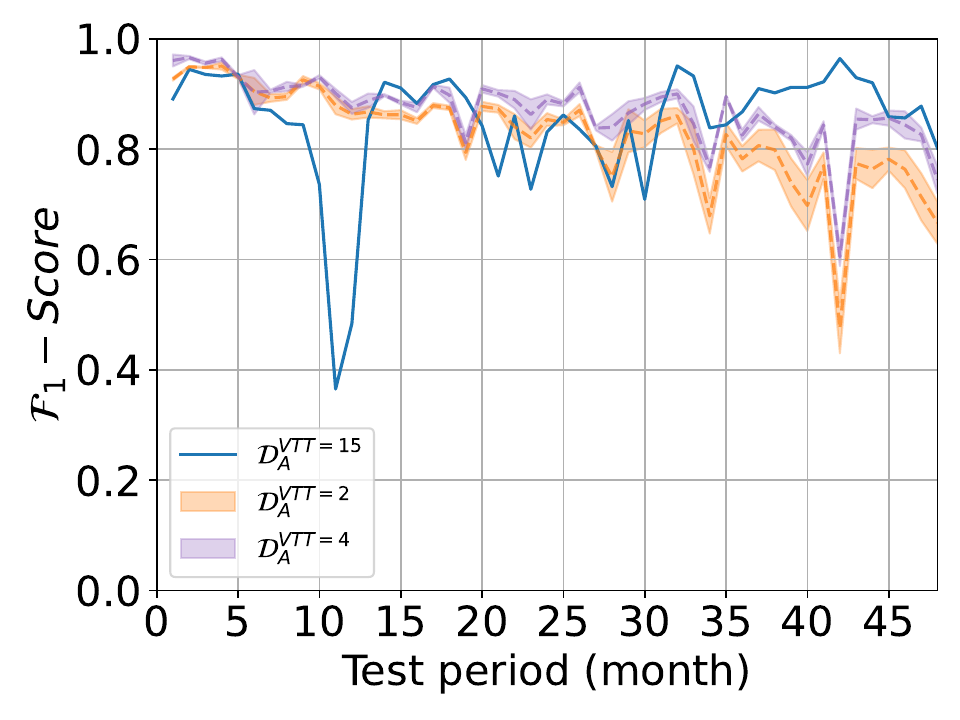}
        \includegraphics[width=0.49\textwidth]{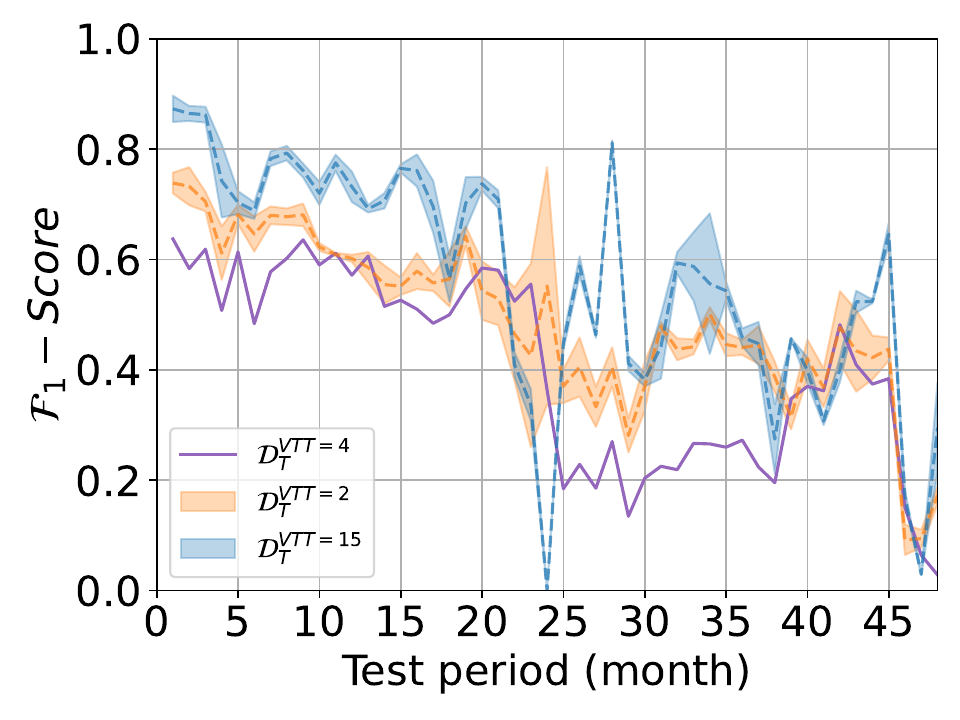}
        \subcaption{\deepdrebin}    
    \end{minipage}
    \hfill
    \begin{minipage}{\linewidth}
        \centering
        \includegraphics[width=0.49\textwidth]{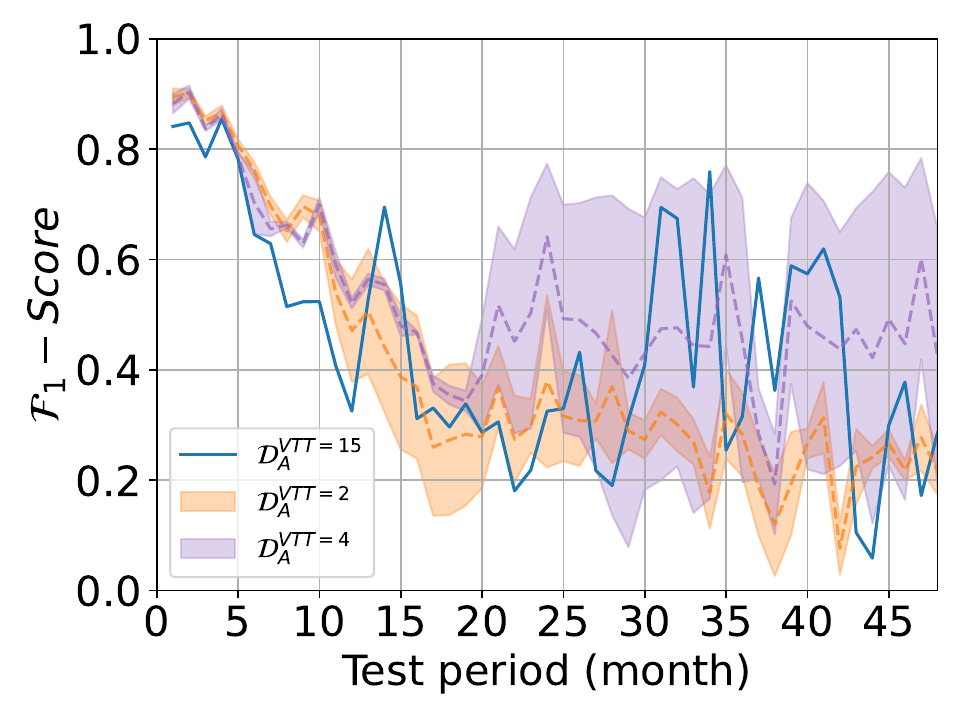}
        \includegraphics[width=0.49\textwidth]{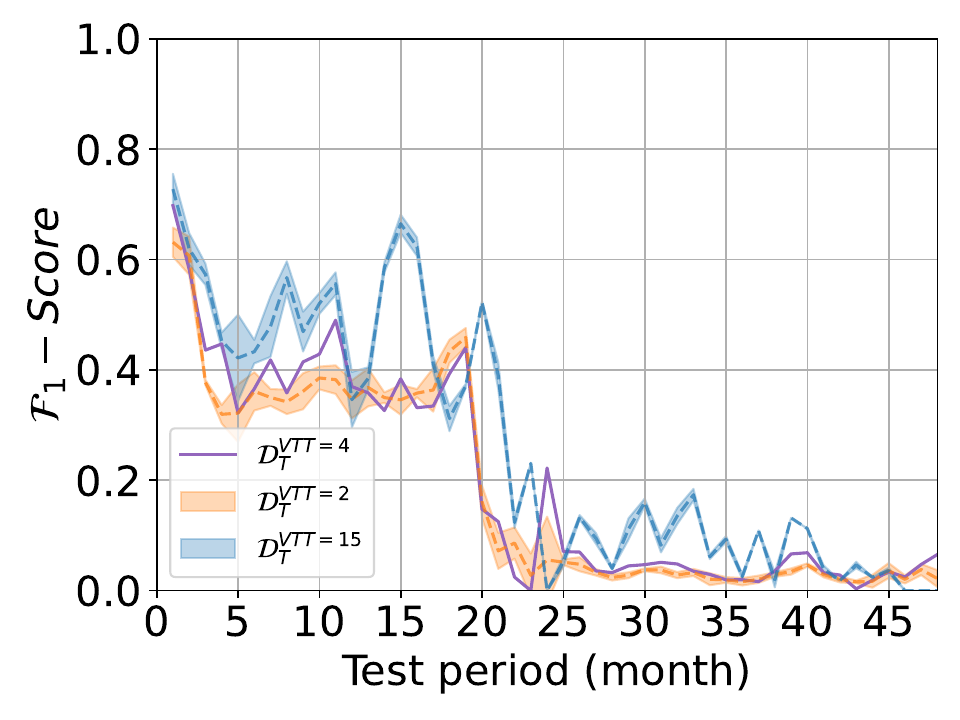}
        \subcaption{\malscan}    
    \end{minipage}
    \hfill
    \begin{minipage}{\linewidth}
        \centering
        \includegraphics[width=0.49\textwidth]{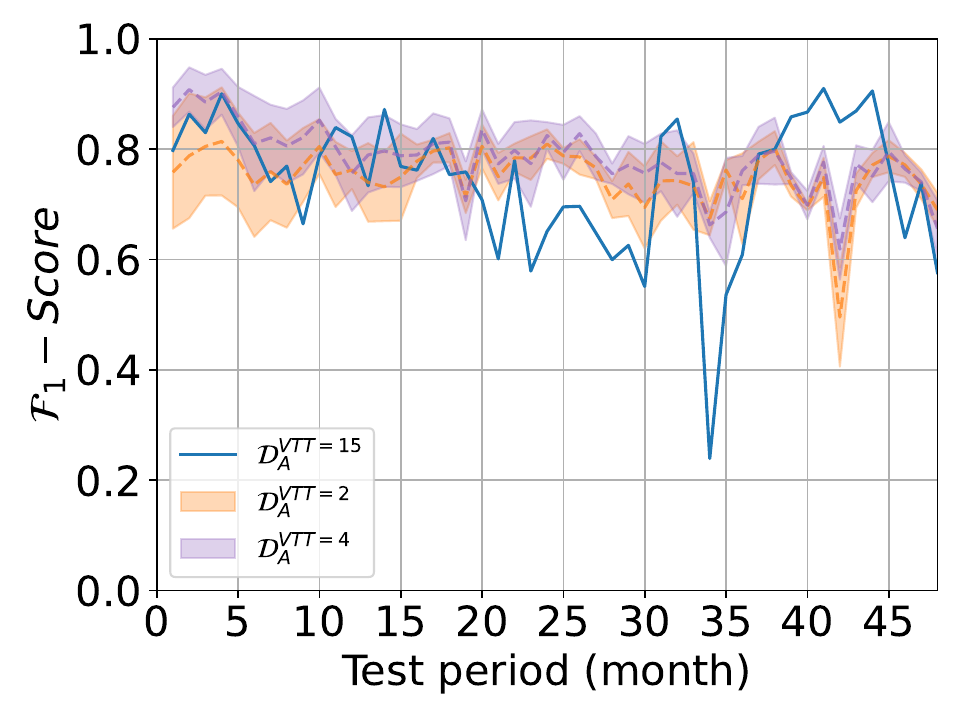}
        \includegraphics[width=0.49\linewidth]{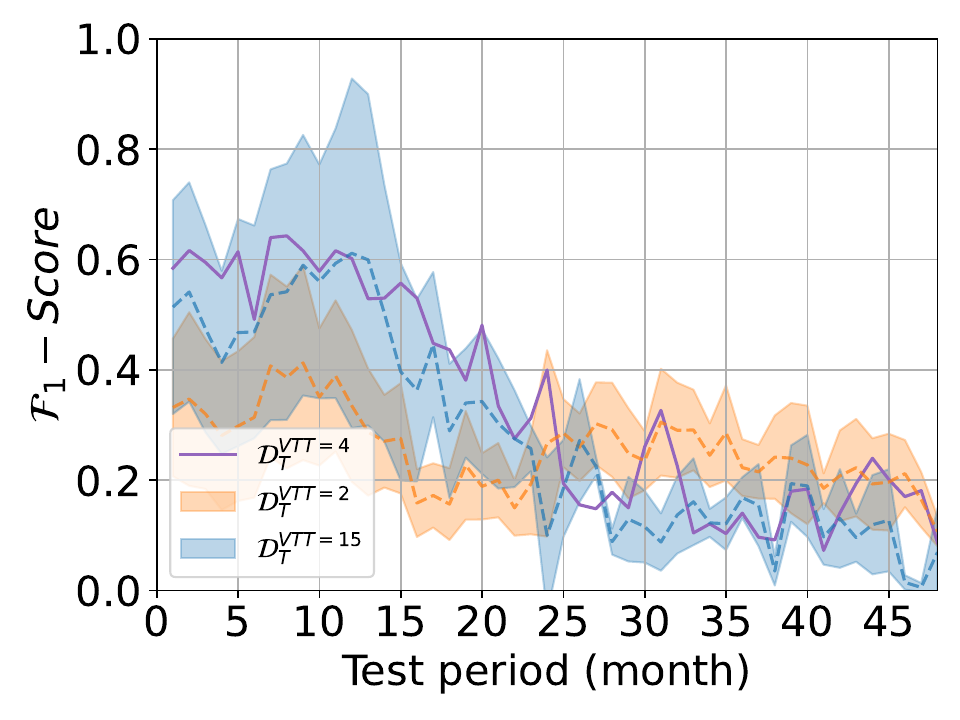}
        \subcaption{\ramda}    
    \end{minipage}
    \hfill
    \begin{minipage}{\linewidth}
        \centering
        \includegraphics[width=0.49\textwidth]{images/vt_exp/apigraph_hcc_diff_new.pdf}
        \includegraphics[width=0.49\linewidth]{images/vt_exp/transcend_hcc_diff_new.pdf}
        \subcaption{\HCC}    
    \end{minipage}
    \caption{\textbf{\fone of classifier for \transcendbold and \apigraphbold sampled with a \ac{vtt}=2, \ac{vtt}=4 and \ac{vtt}=15}}
    \label{fig:vtt_all}
\end{figure}

In \autoref{sec:factor-vtt}, we hypothesized the impact \ac{vtt} has on classification performance. To show this, we sampled both \apigraph and \transcend for \ac{vtt} 2, 4 and 15 and compared it against the original. \apigraph uses a \ac{vtt}=15 and \transcend uses a \ac{vtt}=4. \autoref{sec:appendix:datasets} contains details regarding how we imitated the sampling for both datasets by discarding samples that did not exist in \androzoo when the original dataset was sampled. \autoref{fig:vtt_all}~shows the performance across all 5 different classifiers.

\subsection{Dataset Size}
\label{sec:appendix:dataset-size}
\begin{figure}[t]
    \begin{minipage}{0.49\linewidth}
        \centering
        \includegraphics[width=\linewidth]{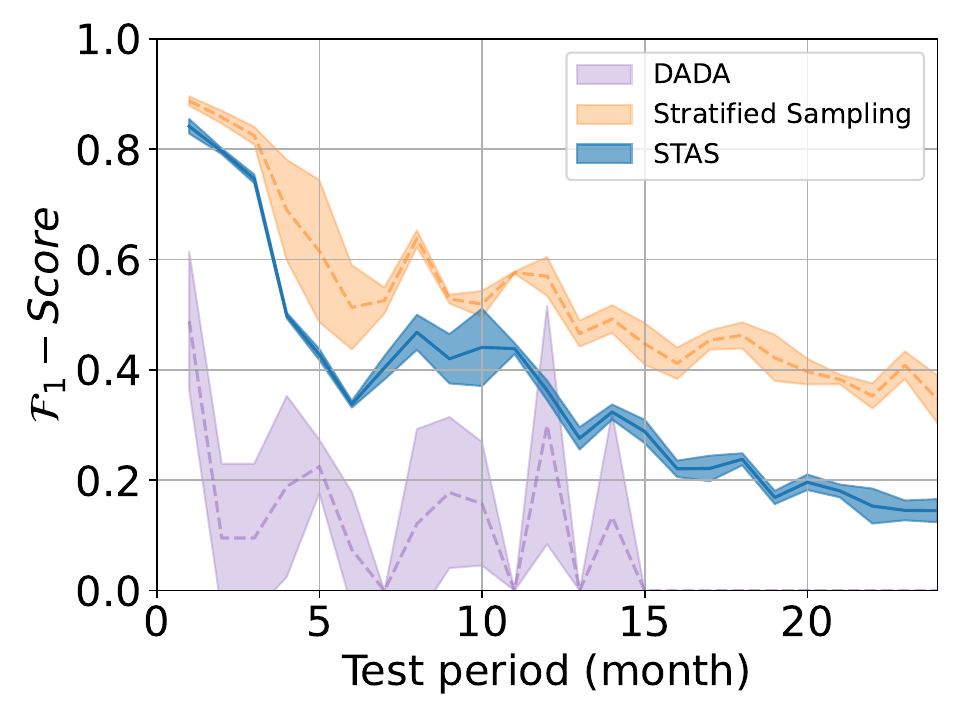}
        \subcaption{\drebin~\cite{arp2014drebin}}
    \end{minipage}
    \begin{minipage}{0.49\linewidth}
        \centering
        \includegraphics[width=\linewidth]{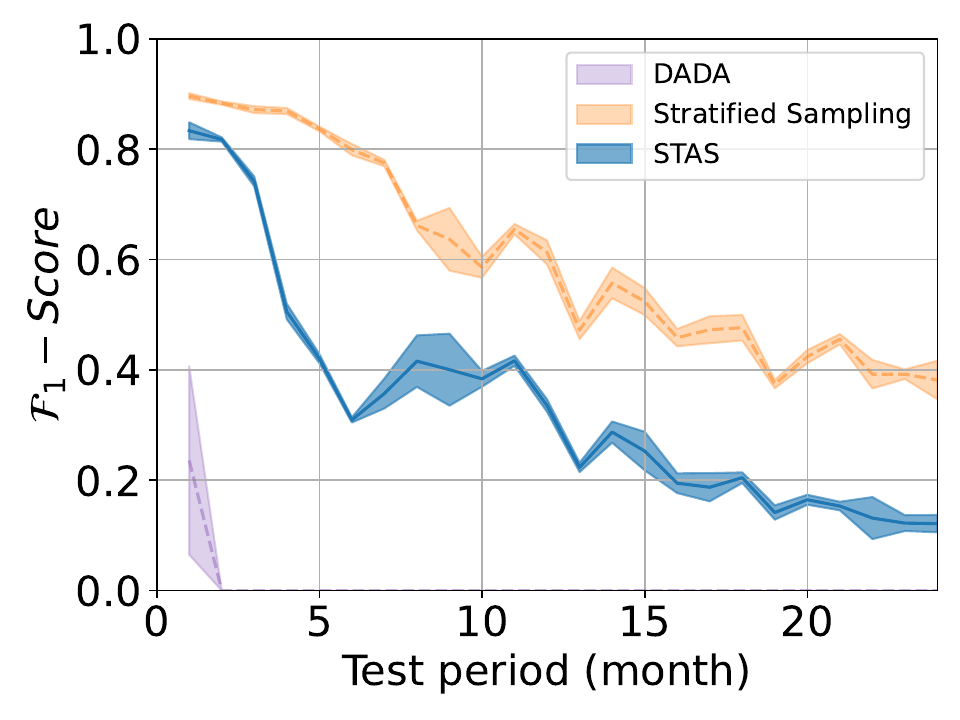}
        \subcaption{\deepdrebin~\cite{deepdrebin}}
    \end{minipage}
    \begin{minipage}{0.49\linewidth}
        \centering
        \includegraphics[width=\linewidth]{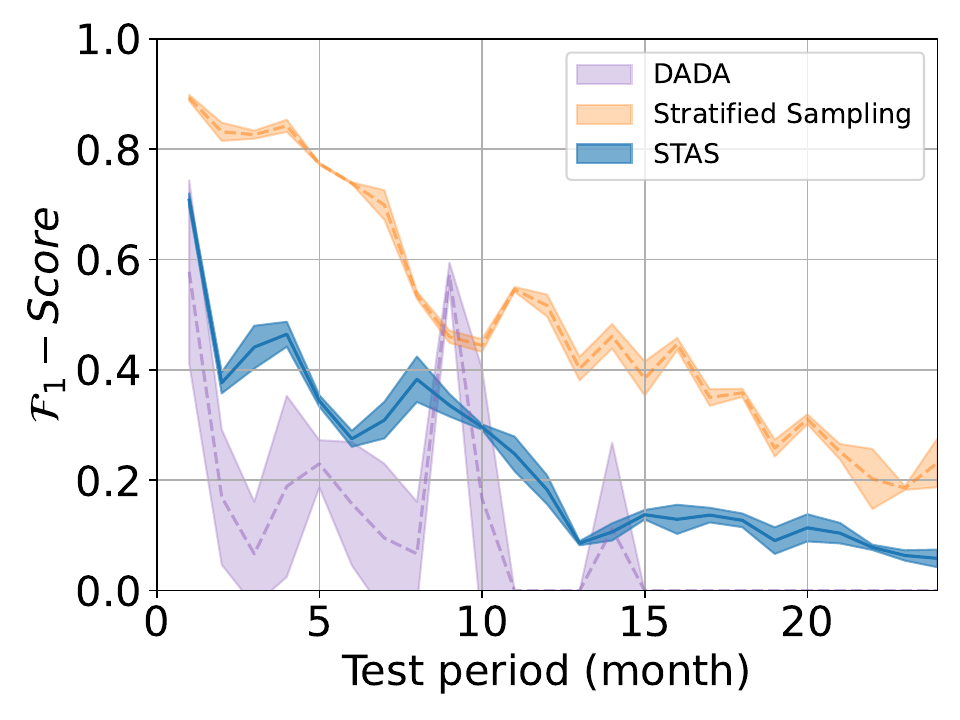}
        \subcaption{\malscan~\cite{malscan_paper}}
    \end{minipage}
    \begin{minipage}{0.49\linewidth}
        \centering
        \includegraphics[width=\linewidth]{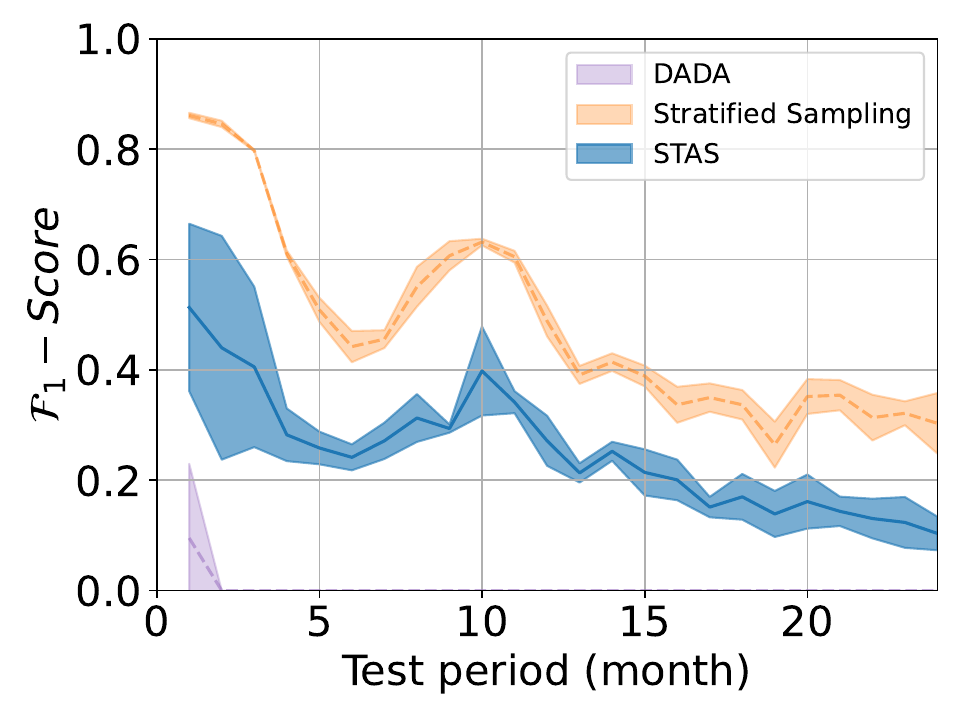}
        \subcaption{\ramda~\cite{ramda}}
    \end{minipage}
    \begin{minipage}{\linewidth}
        \centering
        \includegraphics[width=0.49\linewidth]{images/stability/dada_hypercube3_diff_hcc.pdf}
        \subcaption{\HCC~\cite{chen2023continuous}}
    \end{minipage}
    \caption{\textbf{Comparison of \fone between datasets sampled using \dada versus \stas sampling}. The temporal sample distribution created by \dada is highly inconsistent and, therefore, not appropriate for time-aware evaluations.}
    \label{fig:dada_vs_hypercube3}
\end{figure}

In~\autoref{sec:factor-sampling-size}, we discussed using different sampling strategies for determining the minimum sample size. \autoref{fig:dada_vs_hypercube3}~shows the performance of each strategy for all 5 classifiers.

\subsection{Prevalence Study}
\label{sec:appendix:prevalence}
We begin our analysis by conducting a prevalence study of benchmark datasets used for Android malware detection. Given the vast size of the domain, with over 81,000 results returned when searching ``Android Malware'' on Google Scholar, we narrow our focus. 

We begin with two of the earliest Android malware datasets, Malgenome~\cite{zhou2012dissecting} and Drebin~\cite{arp2014drebin}. Both datasets have over 3,000 citations and were published in top security venues (IEEE S\&P 2012 for Malgenome and NDSS 2014 for Drebin).
In total, we collected 4,995 unique papers that cited either Drebin or Malgenome.

To ensure we only survey relevant high-quality papers, we select works that was published in one of the top four security venues: USENIX Security Symposium, ACM CCS, IEEE S\&P, and NDSS. This helps us filter out papers that do not address security topics or apply machine learning to security-related research problems. After applying the top-4 venue filtering, we ended up with 133 papers.

However, not all notable works was published in the top security venues. To include relevant works that may have been missed from the above criteria, we include the top 100 results of Google Scholar using the search query ``Android Malware Detection'' on July 10th, 2025. 

This forms an initial list of 233 papers, which we review to identify datasets that were used for experimental evaluation. For each paper, we analyzed sections related to dataset curation and evaluation. If a paper did not use a publicly available dataset but collected its own samples, we documented the method of collection. We excluded papers not directly relevant to Android malware detection, including those focused on Windows PE malware, PDF malware, Android UI/UX elements, or privacy policies.

\begin{figure}
    \centering
    \includegraphics[width=1\linewidth]{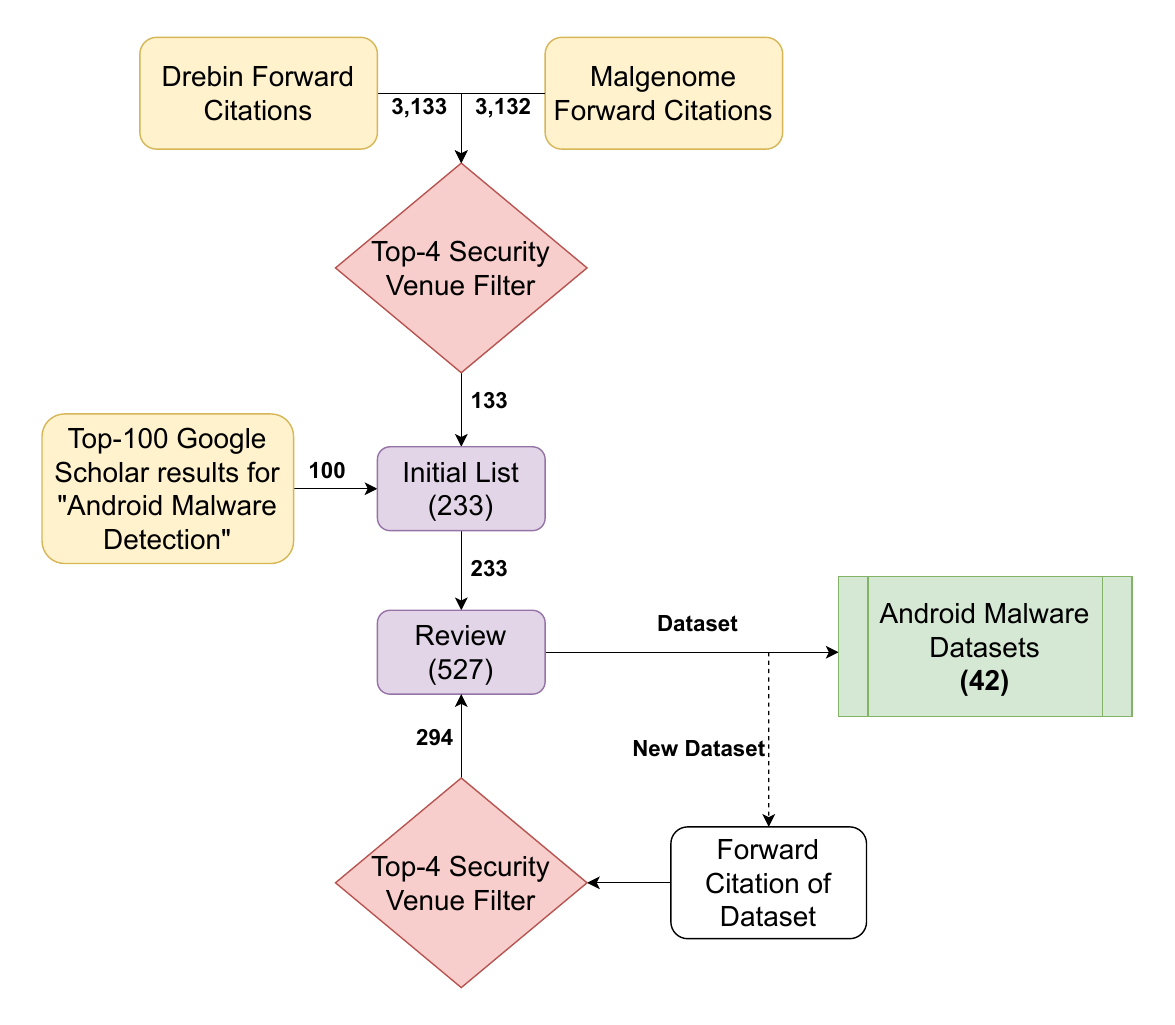}
    \caption{Survey pipeline with queries performed in July 2025. Starting from Drebin and Malgenome, we reviewed papers that were published at one of the top-four security venues. We also reviewed top-100 Google Scholar results for ``Android malware detection.'' We performed forward citation of each new dataset to ensure we did not `miss' any dataset. After reviewing 527 papers, we found 42 Android malware datasets.}
    \label{fig:survey_pipeline}
\end{figure}

From this initial list, we found 35 Android malware datasets used in prior works. To capture the prevalence of these datasets, we performed forward citations (\ie works that cited the dataset) and applied the top four security venue filter to obtain a new list of papers. We review each paper in a similar fashion as above, adding details about newly curated datasets or incrementing the prevalence count of existing ones. We repeat this process until there are no new datasets added. To retrieve forward citations, we used SerpApi~\cite{serpapi}, a Google Scholar API, and verified the publication venues using DBLP~\cite{dblp}, an online bibliographic database for computer science research. Overall, we found 42 unique Android malware datasets. An overview of the survey methodology is presented in \autoref{fig:survey_pipeline}.

We assessed all 42 datasets based on how well they considered each of our factors even if they do not fully satisfy our recommendations. We provide details regarding each factor criterion below.
\paragraph{Timestamp Type}
In~\autoref{sec:factor-timestamps} we show the importance of timestamps when curating a dataset. Therefore, if a dataset explicitly mentions the timestamp used for sampling and evaluation, we consider that they recognized timestamps when sampling. If it only mentioned timestamp for temporal order, we consider it partly satisfied. We noticed some datasets use a combination of other datasets that contain this information, hence mark them as orange. Finally, if a dataset never mentions timestamps and can not be easily inferred by the dataset they chose, then it is marked as red. 

\paragraph{App Markets}
Ideally, datasets should be sampled from a single market/source~(\autoref{sec:factor-app-markets}). Since \androzoo crawls multiple markets, datasets explicitly sampled from a single source or market via \androzoo satisfy our recommendation and are marked green. Datasets that sample from \androzoo without market filtering are marked blue, while those sampling from multiple sources (including combined datasets) are marked red. Finally, gray indicates that market/source information was not provided.

\paragraph{Temporal Luck}
We check if the paper in which each dataset was proposed considers temporal splits during evaluation. If it does, we mark the dataset as green and otherwise as red. Note that we do not consider temporal evaluation or K-fold Cross-validation alone as sufficient.

\paragraph{Dataset Size}
For each dataset, we compute the statistically significant dataset size for each time frame and \ac{vtt} chosen. Here, we only check for statistical significance and do not consider the class ratio where goodware should be nine times the malware amount. If both classes are statistically significant, we mark the dataset as green. If only one class is, we mark it as blue. If neither satisfies the minimum statistical significance, we mark it as red. If a dataset does not mention the dataset size or the minimum is indeterminable, we mark it as gray.

\paragraph{VirusTotal Threshold}
In~\autoref{sec:factor-vtt}, we recommend \ac{vtt}=2. If a dataset uses \ac{vtt}=2, we mark this as green. If a dataset does not use \acl{vtt} and instead opts for pseudo-labels or manual labeling, we mark it as blue, as it does not satisfy our recommendation. If a dataset uses a \ac{vtt} above 2 or mixes multiple different labeling methods (by combining different datasets), we mark this as red. Lastly, if the labeling method was not mentioned, we mark it as gray.

\onecolumn
\begin{table}
\footnotesize
\centering
\rotatebox{90}{%
\scalebox{0.95}{%
    \begin{tabular}{p{3cm}rrp{2cm}ccrrrrp{2cm}}
    \textbf{Dataset} & \# \textbf{Malware} & \# \textbf{Goodware}  &  \textbf{Time Frame} & \textbf{Hashes} &  \textbf{C3}~\cite{pendlebury2019tesseract} & \textbf{Source} & \textbf{Top-4 Prevalence}&\textbf{Prevalence}&\textbf{Venue} \\
    \toprule
         Malgenome~\cite{zhou2012dissecting} & 1,260 & 0 & 2010-2011 & \emptycircle & \emptycircle & & 3&20 & S\&P 2012 \\
         Liu et al.~\cite{liu2014two} & 1,536 & 28,548 & 2012 & \emptycircle & \fullcircle & \ding{58}, \cite{zhou2012dissecting} & 0 & 1 & MobileCloud 2014 \\
         Andrubis~\cite{lindorfer2014andrubis} & 41.14\% & 27.90\% & 2010-2014 & \emptycircle & \halfcircle & $GP$,\ding{58},\cite{zhou2012dissecting, arp2014drebin} & 0 & 3 & ESORICS 2014 \\
         DroidDolphin~\cite{wu2014droiddolphin} & 32,000 & 32,000 & - & \emptycircle & \halfcircle & $GP$ & 0 & 1 & RACS 2014 \\
         \textbf{Drebin}~\cite{arp2014drebin} & 5,560 & 123,453 & 2010-2012 & \fullcircle & \fullcircle & \cite{zhou2012dissecting}, $AZ$, $GP$ & 9 & 46 & NDSS 2014\\
         Triggerscope~\cite{fratantonio2016triggerscope} & - & 9,582 & - & \emptycircle & \emptycircle & $GP$ & 1 & 1 & S\&P 2016 \\
         ICCDetector~\cite{xu2016iccdetector} & 5,264 & 12,026 & 2012 & \emptycircle & \halfcircle & $GP$, \cite{arp2014drebin} & 0 & 1 & TIFS 2016\\
         Anastasia~\cite{fereidooni2016anastasia} & 18,677 & 11,187 & 2009-2015 & \emptycircle & \halfcircle & \cite{zhou2012dissecting,arp2014drebin}, $VT$ & 0 & 1 & NTMS 2016\\
         Droidnative~\cite{alam2017droidnative} & 5,490 & 3,732 & - & \emptycircle & \halfcircle & \cite{arp2014drebin, zhou2012dissecting} & 0 & 1 & Computers \& Security 2017 \\
         Piggybacking~\cite{piggybacking} & 1,497 & 0 & 2009-2014 & \fullcircle & \emptycircle & $GP$, \ding{58}, \cite{zhou2012dissecting} & 0 & 1 & TIFS 2017 \\
         Shen et al.~\cite{shen2018android} & 4,699 & 3,899 & 2014-2016 & \emptycircle & \halfcircle & $GP$, \cite{zhou2012dissecting} & 0 & 1 & ICDCS 2017 \\
         AMD~\cite{wei2017deep} & 24,553 & 0 & 2010-2016 & \emptycircle & \emptycircle & $GP$, $VS$, \ding{58} & 7 & 29 & DIMVA 2017 \\
         MamaDroid~\cite{mariconti2017mamadroid} & 35,500 & 8,500 & 2010-2016 & \halfcircle & \fullcircle & \cite{arp2014drebin}, $VS$, $GP$ & 0 & 2 & NDSS 2017\\
         Faldroid~\cite{faldroid} & 8,407 & 6,593 & - & \fullcircle & \halfcircle & $VS$ & 1 & 1 & TIFS 2018\\
         EnDroid~\cite{feng2018novel} & 55,213 & 58,806 & 2015-2018 & \emptycircle & \halfcircle & $GP$, $AZ$, \cite{arp2014drebin} &0 & 1 &  IEEE Access 2018\\
         Flowcog~\cite{pan2018flowcog} & 1,500 & 4,500 & - & \fullcircle & \halfcircle & $GP$, \cite{arp2014drebin} & 1 & 1 & USENIX 2018 \\
         Maldozer~\cite{karbab2018maldozer} & 20,089 & 37,627 & - & \emptycircle & \halfcircle & $VS$, \cite{zhou2012dissecting,arp2014drebin,contagio} & 0 & 1 & Digital Investigation 2018\\
         CICAndMal2017~\cite{cicandmal2017} & 429 & 5,065 & 2015-2017 & \emptycircle & \halfcircle & $VT$, \cite{contagio}, \cite{gonzalez2015droidkin, abdul2015android, lashkari2017towards} & 0 & 4 & ICCST 2018\\
         Hasegawa et al.~\cite{hasegawa2018one} & 5,000 & 2,000 & - & \emptycircle & \halfcircle & \ding{58}, \cite{zhou2012dissecting,arp2014drebin} & 0 & 1 & CSPA 2018 \\
         Li et al.~\cite{li2018significant} & 54,694 & 310,926 & 2012-2014 & \halfcircle & \fullcircle & $GP$, \ding{58} & 0 & 1 & IEEE Industrial Inf. 2018\\
         Marvin~\cite{lindorfer2015marvin} & $\approx$15,000 & $\approx$120,000 & 2012-2014 & \halfcircle & \fullcircle & $GP$, \ding{58} & 1 & 2 & ASE 2019 \\
         Malscan~\cite{malscan_paper} & 15,430 & 15,285 & 2011-2018 & \fullcircle & \halfcircle & $AZ$ & 1 & 1 & ASE 2019 \\
         \textbf{Tesseract}~\cite{pendlebury2019tesseract} & 12,753 & 116,993 & 2014-2016 & \fullcircle & \fullcircle & $AZ$ & 3 & 3 & USENIX 2019 \\
         DroidCat~\cite{cai2018droidcat} & 135 & 136 & 2015 & \halfcircle & \halfcircle & $GP$, \cite{zhou2012dissecting} & 0 & 1 &TIFS 2019 \\
         Omnidroid~\cite{martin2019android} & 11,000 & 11,000 & - & \fullcircle & \halfcircle & $AZ$ & 0 & 1 &Information Fusion 2019 \\
         \textbf{Pierazzi et al.}~\cite{pierazzi2020intriguing} & 17,635 & 152,632 & 2017-2018 & \fullcircle & \fullcircle & $AZ$ & 2 & 2 & S\&P 2020 \\
         CICMalDroid2020~\cite{mahdavifar2020dynamic} & 17,341 & 0 & 2018 & \emptycircle & \emptycircle & $VT$, \cite{wei2017deep, contagio, karbab2018maldozer, abdul2015android, stakhanova2016empirical} & 0 & 3 & DASC 2020\\
         \textbf{APIGraph}~\cite{apigraph_paper} & 32,089 & 290,505 & 2012-2018 & \fullcircle & \fullcircle & $VS$, $VT$, \cite{wei2017deep}, $AZ$ & 2 & 2 &CCS 2020 \\
         Dhalaria et al.~\cite{dhalaria2021hybrid} & 1,747 & 1,800 & - & \fullcircle & \halfcircle & $VT$, \ding{58} & 0 & 1 & IMAI 2020 \\
         Kronodroid~\cite{guerra2021kronodroid} & 28,745 & 35,256 & 2008-2020 & \fullcircle & \halfcircle & \cite{deepdrebin, wei2017deep} $VT$, $VS$ & 0 & 2 & Elsevier CoSe 2021 \\
         AndroCT~\cite{li2021androct} & 17,679 & 18,277 & 2019-2019 & \fullcircle & \halfcircle & $AZ$, $VS$, $GP$ & 0 & 1 & MSR 2021\\
         VTAz~\cite{frenklach2021android} & 5,016 & 4,987 & 2017-2020 & \halfcircle & \halfcircle & $VT$, $AZ$ &  0 & 1 & Computer Security 2021\\
         Sun et al.~\cite{sun2021detecting} & 12,685 & 49,045 & 2014 & \emptycircle & \fullcircle & $GP$, \cite{wei2017deep} & 0 & 1 & TMIS 2022 \\
         DroidMalware Detector~\cite{kabakus2022droidmalwaredetector} & 7,752 & 6,634 & - & \emptycircle & \halfcircle & $AZ$, $VS$, \cite{arp2014drebin, chen2016towards} & 0 & 1 & \makecell[r]{Expert Systems\\with Applications 2022} \\
         Malradar~\cite{wang2022malradar} & 4,534 & 0 & 2014-2021 & \fullcircle & \emptycircle & & 0 & 1 & ACM MACS 2022\\
         Madroid~\cite{wang2022malwhiteout} & 35,121 & 39,571 & 2010-2023 & \fullcircle & \halfcircle & $AZ$, \cite{zhou2012dissecting,malscan_paper, allix2016empirical} & 0 & 2 & ASE 2022 \\
         AndroOBFS~\cite{kumar2022androobfs} & 16,279 & 0 & 2018-2020 & \fullcircle & \emptycircle & $AZ$, $VS$ & 0 & 1 & MSR 2022\\
         \textbf{Transcendent}~\cite{barbero2022transcending} & 26,387 & 232,843 & 2014-2018 & \fullcircle & \fullcircle & $AZ$ & 2 & 2 & S\&P 2022 \\
         Finer~\cite{he2023finer} & 4,742 & 12,807 & 2017-2019 & \fullcircle & \halfcircle & - & 1 & 1 & CCS 2023\\
         \textbf{Androzoo}~\cite{chen2023continuous} & 10,200 & 89,853 & 2019-2021 & \halfcircle & \fullcircle & AZ & 1 & 1 & USENIX 2023\\
         Heng et al.~\cite{li2023black} & 22,975 & 21,399 & - & \halfcircle & \halfcircle & $AZ$, \cite{faldroid,arp2014drebin} & 1 & 1 & USENIX 2023\\
         Jigsaw~\cite{yang2023jigsaw} & 14,775 & 134,759 & 2015-2016 & \halfcircle & \fullcircle & $AZ$ & 0 & 1 & S\&P 2023 \\
         Maldroid-DS~\cite{almomani2024maloid} & 47,971 & 0 & 2010-2024 & \fullcircle & \emptycircle & \cite{wei2017deep, arp2014drebin, keyes2021entroplyzer, cicandmal2017}, $VS$ & 0 & 1 & IEEE Access 2024\\
    \bottomrule
    \end{tabular}
}
}
    \caption{Breakdown of existing Android malware datasets discovered in our survey, arranged by publication date. $AZ$~samples crawled from Androzoo repository~\cite{allix2016androzoo}, $GP$~samples crawled directly from \googleplay, $VS$~samples from VirusShare, \ding{58}~samples crawled from other sources.}
    \label{tab:dataset_used}
\end{table}

\end{document}